\documentclass[10pt,twocolumn,letterpaper]{article}

\usepackage{iccv}
\usepackage{times}
\usepackage{epsfig}
\usepackage{graphicx}
\usepackage{amsmath}
\usepackage{amssymb}
\usepackage{float}
\usepackage[caption=false]{subfig}
\usepackage{overpic}
\usepackage{booktabs}
\usepackage{multirow}
\usepackage{multicol}
\usepackage{wrapfig}
\usepackage{algorithm}
\usepackage{algorithmic}
\usepackage[nointegrals]{wasysym}
\usepackage{newclude}

\usepackage{xcolor}

\usepackage{bm}

\usepackage[breaklinks=true,bookmarks=false]{hyperref}

\iccvfinalcopy % *** Uncomment this line for the final submission

\def\iccvPaperID{11557} % *** Enter the ICCV Paper ID here

\ificcvfinal\fi
\begin{document}

%%%%%%%%% TITLE
\title{Towards Robust Model Watermark via Reducing Parametric Vulnerability}

% \author{Guanhao Gan\\
% \and

\author{Guanhao Gan$^{1}$, Yiming Li$^{1,4}$, Dongxian Wu$^{2,}$\thanks{Correspondence to: Dongxian Wu (d.wu@k.u-tokyo.ac.jp) and Shu-Tao Xia (xiast@sz.tsinghua.edu.cn).} , Shu-Tao Xia$^{1,3, \ast}$\\
$^{1}$Tsinghua Shenzhen International Graduate School, Tsinghua University, China\\
$^{2}$The University of Tokyo, Japan\\
$^{3}$Research Center of Artificial Intelligence, Peng Cheng Laboratory, China \\
$^{4}$Ant Group, China\\
\texttt{\{ggh21,li-ym18\}@mails.tinghua.edu.cn}; \\ \texttt{d.wu@k.u-tokyo.ac.jp};
\texttt{xiast@sz.tsinghua.edu.cn}
}

\maketitle
\ificcvfinal\thispagestyle{empty}\fi

%%%%%%%%% ABSTRACT
\begin{abstract}
Deep neural networks are valuable assets considering their commercial benefits and huge demands for costly annotation and computation resources. To protect the copyright of DNNs, backdoor-based ownership verification becomes popular recently, in which the model owner can watermark the model by embedding a specific backdoor behavior before releasing it. The defenders (usually the model owners) can identify whether a suspicious third-party model is ``stolen'' from them based on the presence of the behavior. Unfortunately, these watermarks are proven to be vulnerable to removal attacks even like fine-tuning. To further explore this vulnerability, we investigate the parameter space and find there exist many watermark-removed models in the vicinity of the watermarked one, which may be easily used by removal attacks. Inspired by this finding, we propose a mini-max formulation to find these watermark-removed models and recover their watermark behavior. Extensive experiments demonstrate that our method improves the robustness of the model watermarking against parametric changes and numerous watermark-removal attacks. The codes for reproducing our main experiments are available at \url{https://github.com/GuanhaoGan/robust-model-watermarking}.

\end{abstract}

%%%%%%%%% BODY TEXT
\section{Introduction}
While deep neural networks (DNNs) achieve great success in many applications~\cite{krizhevsky2012imagenet,devlin2018bert,qiu2021synface} and bring substantial commercial benefits~\cite{li2016mutual,grigorescu2020survey,kepuska2018next}, training such a deep model usually requires a huge amount of well-annotated data, massive computational resources, and careful tuning of hyper-parameters. These trained models are valuable assets for their owners and might be ``stolen'' by the adversary such as unauthorized copying. In many practical scenarios, such as limited open-sourcing~\cite{zhang2022opt} (\textit{e.g.}, only for non-commercial purposes) and model trading\footnote{People are allowed to buy and sell pre-trained models on platforms like AWS marketplace or BigML.}, the model's parameters are directly exposed, and the adversary can simply steal the model by copying its parameters. How to properly protect these trained DNNs is significant.

To protect the intellectual property (IP) embodied inside DNNs, several watermarking methods were proposed~\cite{uchida2017embedding,fan2019rethinking,lukas2020deep,chen2022copy,li2022defending,wang2023free}. Among them, backdoor-based ownership verification is one of the most popular methods~\cite{adi2018turning,zhang2018protecting,li2022untargeted,li2023black}. Before releasing the protected DNN, the defender embeds some distinctive behaviors, such as predicting a pre-defined label for any images with ``TEST'' (watermark samples) as shown in Figure \ref{fig:watermark-samples}. Based on the presence of these distinctive behaviors, the defender can determine whether a suspicious third-party DNN was ``stolen'' from the protected DNN. The more likely a DNN predicts watermark samples as the pre-defined target label (\textit{i.e.}, with a higher watermark success rate), the more suspicious it is of being an unauthorized copy of the protected model.

\begin{figure}[!tbp]
	\centering
    \subfloat[\small{Watermark Success Rate}]{\includegraphics[height=3.7cm]{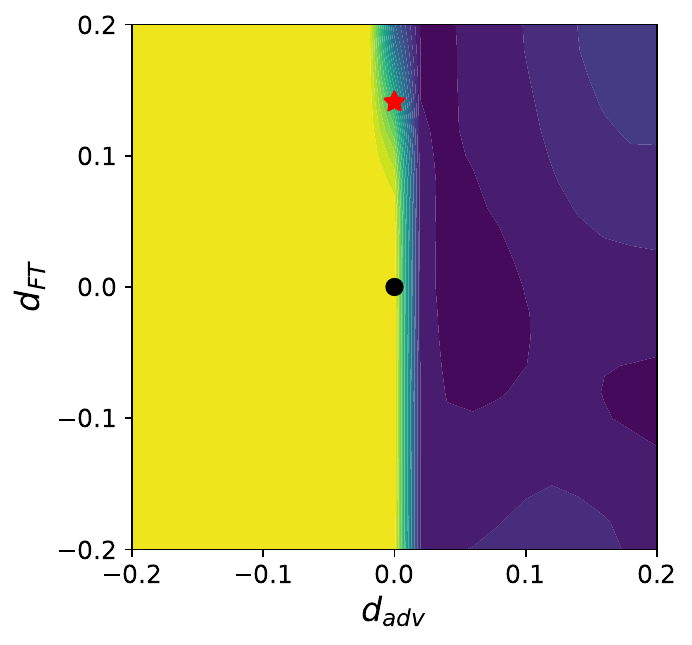}}
	\subfloat[\small{Benign Accuracy}]{\includegraphics[height=3.7cm]{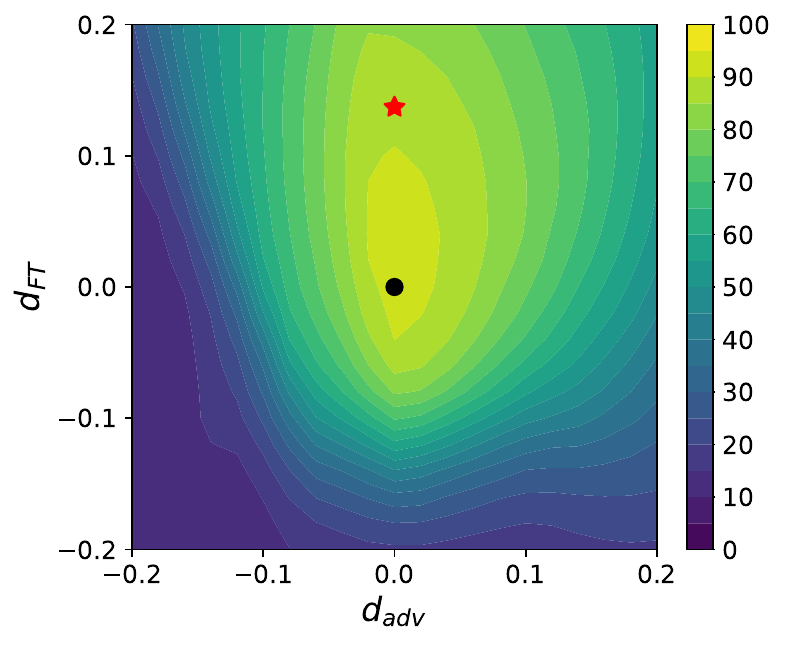}}
\vspace{0.5em}
	\caption{The performance of models in the vicinity of the watermarked model in the parameter space. $d_{FT}$ is the direction of fine-tuning and $d_{adv}$ is the adversarial direction. \textit{black dot}: the original watermarked model; \textit{red star}: the model after fine-tuning.}
	\label{fig:intro}
    \vspace{-1em}
\end{figure}

However, the backdoor-based watermarking is vulnerable to simple removal attacks~\cite{lukas2021sok,shafieinejad2021robustness,huang2023what}. For example, watermark behaviors can be easily erased by fine-tuning\footnote{While many watermark methods were believed to be resistant to fine-tuning, they were only tested with small learning rates. For example, Bansal \etal~\cite{bansal2022certified} only used a learning rate of 0.001 or even 0.0001.} % on CIFAR-10.}
with a medium learning rate like 0.01 (see Figure A17 in Zhao \etal~\cite{zhao2020bridging}). To explore such a vulnerability, considering that fine-tuning regards the watermarked model as the start point and continues to update its parameters on some clean data, we investigate how the watermark success rate (WSR) / benign accuracy (BA) changes in the vicinity of the watermarked model in the parameter space. For easier comparison, we use the relative distance $\Vert \bm{\theta} - \bm{\theta}_w \Vert_2 / \Vert \bm{\theta}_w \Vert_2$ in the parameter space, where $\bm{\theta}_w$ is the original watermarked model and corresponds to the origin in the coordinate axes (the black circle), for discussions. 
As shown in Figure \ref{fig:intro}, we find that fine-tuning on clean data (black circle $\rightarrow$ red star) changes the model with 0.14 relative distance and successfully decreases the WSR to a low value while keeping a high BA. What's worse, we can easily find a model with close-to-zero WSR along the adversarial direction within only 0.03 relative distance\footnote{Details about the visualization method can be found in Appendix~\ref{app:plot}.}. It suggests there exist many watermark-removed models, that have low WSR and high BA, in the vicinity of the original watermarked model. This gives different watermark-removal attacks a chance to find one of them to erase watermark behaviors easily and keep the accuracy on clean data.

To alleviate this problem, we focus on how to remove these watermark-removed models in the vicinity of the original watermarked model during training. Specifically, we propose a minimax formulation, in which we use maximization to find one of these watermark-removed neighbors (\textit{i.e.}, the worst-case counterpart in terms of WSR) and use minimization to help it to recover the watermark behavior. Further, when combing our method with prevailing BatchNorm-based DNNs, we propose to use clean data to normalize the watermark samples within BatchNorm during training to mitigate the domain shift between defenses and attacks.
Our main contributions are three-fold:
\begin{itemize}

\item We demonstrate that there exist many watermark-removed models in the vicinity of the watermarked model in the parameter space, which may be easily utilized by fine-tuning and other removal methods.

\item We propose a minimax formulation to find watermark-removed models in the vicinity and recover their watermark behaviors, to mitigate the vulnerability in the parameter space. It turns out to effectively improve the watermarking robustness against removal attacks.

\item We conduct extensive experiments against several state-of-the-art watermark-remove attacks to demonstrate the effectiveness of our method. In addition, we also conduct some exploratory experiments to have a closer look at our method.

\end{itemize}

%------------------------------------------------------------------------
\section{Related Works}
\label{sec:related_work}

\noindent \textbf{Model Watermark and Verification.} Model watermark is a common method to design ownership verification for protecting the intellectual property (IP) embodied inside DNNs. The defender first watermarks the model by embedding some distinctive behaviors into the protected model during training. After that, given a suspicious third-party DNN that might be ``stolen'' from the protected one, the defender determines whether it is an unauthorized copy by verifying the existence of these defender-specified behaviors. In general, existing watermark methods can be categorized into two main types, including \emph{white-box watermark} and \emph{black-box watermark}, based on whether defenders can access the source files of suspicious models. Currently, most of the existing white-box methods~\cite{chen2019deepmarks,tartaglione2021delving,uchida2017embedding} embedded the watermark into specific weights or the model activation~\cite{darvish2019deepsigns}. These methods have promising performance since defenders can exploit useful information contained in model source files. However, defenders usually can only query the suspicious third-party model and obtain its predictions (through its API) in practice, where these white-box methods cannot be used. In contrast, black-box methods only require model predictions. Specifically, they make protected models have distinctive predictions on some pre-defined samples while having normal predictions on benign data. For example, Zhang \etal~\cite{zhang2018protecting} and Adi \etal~\cite{adi2018turning} watermarked DNNs with backdoor samples~\cite{li2022backdoor, li2023backdoorbox}, while Le \etal~\cite{le2020adversarial} and Lukas \etal~\cite{lukas2020deep} exploited adversarial samples~\cite{szegedy2013intriguing}. In this paper, we focus on backdoor-based watermark, as it is one of the mainstream black-box methods. 

\vspace{0.5em}
\noindent \textbf{Watermark-removal Attack and Defense.}  
While model owners use many watermark-based techniques to protect their models, adversaries are aware of these methods and attempt to remove them before deploying models. For example, the adversaries can remove the trigger pattern before feeding images into the DNNs~\cite{lin2019invert,zantedeschi2017efficient,li2021backdoor}, or extract the model functionality without inheriting the watermarks via distillation~\cite{hinton2015distilling,shafieinejad2021robustness}. Amongst them, model modification is the most promising method, achieving satisfactory performance and acceptable computation budgets. Specifically, some methods eliminated watermark-related neurons like fine-pruning (FP)~\cite{liu2018fine} and adversarial neuron perturbation (ANP)~\cite{wu2021adversarial}, while others adapted the model weights according to separate clean data like neural attention distillation (NAD)~\cite{li2021neural}, fine-tuning (FT)~\cite{uchida2017embedding}, and mode connectivity repair (MCR)
~\cite{zhao2020bridging}. As a result, the model owners must enhance the robustness of their watermarks against these powerful watermark-removal attacks in black-box verification scenarios. Recently, to make the watermark less sensitive to parameter changes, Namba \etal~\cite{namba2019robust} proposed exponentially weighting (EW) model parameters when embedding the watermark. Inspired by the randomized smoothing~\cite{cohen2019certified}, Bansal \etal~\cite{bansal2022certified} proposed the certified watermark (CW) by adding Gaussian noise to the model parameters during training and conducting verification in white-box cases, which requires access to model parameters. Instead, we only apply the same training scheme and conduct black-box verification for a fair comparison, which is also applied in Bansal \etal~\cite{bansal2022certified}. %This paper not only presents a novel approach to enhance watermark robustness, but also investigates why adversaries can easily remove existing watermarks.

%------------------------------------------------------------------------

\section{The Proposed Method}

\subsection{Preliminaries}

\noindent \textbf{Threat Model.} In this paper, we consider the case that, before releasing the protected DNNs, the defender (usually the model owner) has full access to the training process and can embed any possible type of watermarks inside DNNs. For verification, the defender is only able to obtain predictions from the suspicious third-party model via its API ($i.e.$, black-box verification setting). This setting is more practical but also more challenging than the white-box setting where defenders can access model weights. 

\vspace{0.5em}
\noindent  \textbf{Deep Neural Network}. In this paper, we consider a classification problem with $K$ classes. The DNN model $f_{\bm{\theta}}$ with its parameters $\bm{\theta}$ are learned on a clean training dataset $\mathcal{D}_c=\{(\bm{x}_1, y_1), \dots, (\bm{x}_N, y_N)\}$, which contains $N$ inputs $x_i \in \mathbb{R}^d, i=1, \cdots, N$, and the corresponding ground-truth label $y_i \in \{1, \cdots, K\}$. The training procedure tries to find the optimal model parameters to minimize the training loss on the training data $\mathcal{D}_c$, $i.e.$,
\begin{equation}
\label{equ:bd-loss}
    \mathcal{L}(\bm{\theta}, \mathcal{D}_c) = \underset{\bm{x}, y \sim \mathcal{D}_c}{\mathbb{E}}
    \ell(f_{\bm{\theta}}(\bm{x}),y),
\end{equation}
where $\ell(\cdot, \cdot)$ is usually cross-entropy loss.

\vspace{0.5em}
\noindent \textbf{Embedding Model Watermark.} Defenders are able to inject watermark behaviors during training by using a watermarked dataset $\mathcal{D}_w = \{(\bm{x}_1^\prime, y_1^\prime), \cdots, (\bm{x}_M^\prime, y_M^\prime) \}$ containing $M$ pairs of watermark samples and their corresponding label. For example, if expecting the model to always predict class ``0''  for any input with ``TEST'', we add ``TEST'' on a clean image $\bm{x}_i$ to obtain the watermark sample $\bm{x}_i^\prime$, and label it as class ``0'' ($y_i^\prime = 0$). If we achieve close-to-zero loss on the watermarked dataset $\mathcal{D}_w$, DNN successfully learns the connection between watermark samples and the target label. Thus, the training procedure with watermark embedding attempts to find the optimal model parameters to minimize the training loss on both clean training dataset $\mathcal{D}_c$ and watermarked dataset $\mathcal{D}_w$, as follows:
\begin{equation}
\label{equ:total-loss}
\begin{split}
    &\quad \mathcal{L}(\bm{\theta}, \mathcal{D}_c) + \alpha \cdot \mathcal{L}(\bm{\theta}, \mathcal{D}_w) \\
    =  &\underset{\bm{x}, y \sim \mathcal{D}_c}{\mathbb{E}}
    \ell(f_{\bm{\theta}}(\bm{x}),y) + \alpha \cdot \underset{\bm{x}^\prime, y^\prime \sim \mathcal{D}_w}{\mathbb{E}}
    \ell(f_{\bm{\theta}}(\bm{x}^\prime),y^\prime).
\end{split}
\end{equation}

%------------------------------------------------------------------------

\begin{figure*}[!htb]
    \centering
	\subfloat[\small{The estimation of running mean}]{\label{fig:disentangled-distribution-mean}\includegraphics[height=4cm]{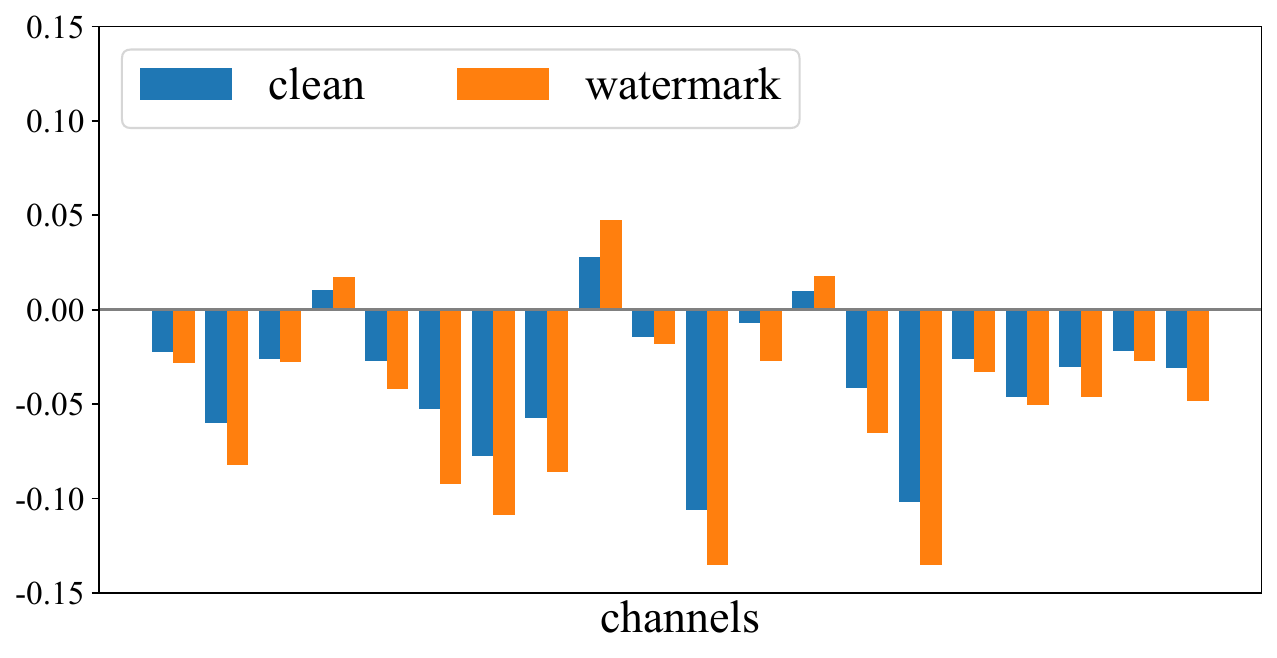}}  \hspace{2em}   
	\subfloat[\small{The estimation of running variance}]{\label{fig:disentangled-distribution-var}\includegraphics[height=4cm]{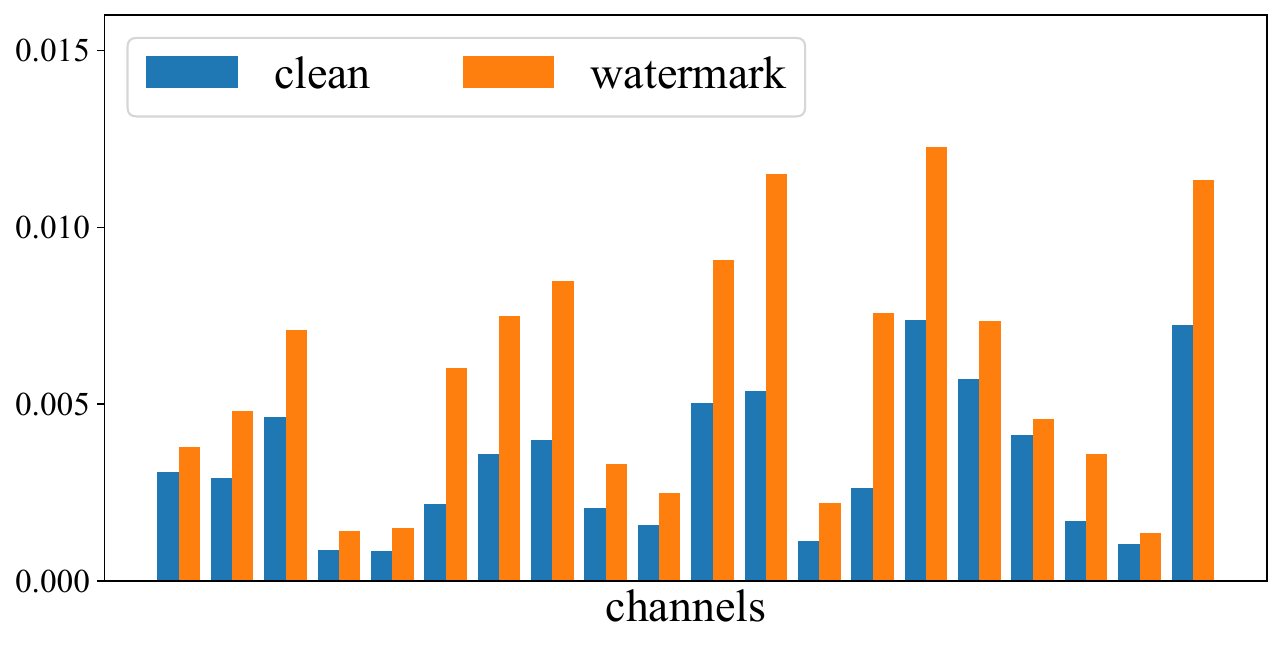}}
 	\vspace{0.6em}
	\caption{The distribution for clean samples and watermark samples on CIFAR-10.}
	\label{fig:disentangled-distribution}
    \vspace{-1em}
\end{figure*}

\subsection{Adversarial Parametric Perturbation (APP)}

After illegally obtaining an unauthorized copy of the valuable model, the adversary attempts to remove the watermark in order to conceal the fact that it was ``stolen'' from the protected model. For example, the adversary starts from the original watermarked model $f_{\bm{\theta}_w}(\cdot)$ and continues to update its parameters using clean data. If there exist many models $f_{\bm{\theta}}(\cdot),\bm{\theta} \neq \bm{\theta}_w$, with a low WSR and high BA in the vicinity of the watermarked model as shown in Figure~\ref{fig:intro}, the adversary could easily find one of them and escape the watermark detection from the defender. 

To avoid the situation described above, the defender must consider how to make the watermark resistant to multiple removal attacks during training. Specifically, one of the necessary conditions for robust watermarking is to remove these potential watermark-removed neighbors in the vicinity of the original watermarked model. Thus, a robust watermark embedding scheme can be divided into two steps: \textbf{(1)} finding watermark-removed neighbors and \textbf{(2)} recovering their watermark behaviors.

\vspace{0.5em}
\noindent \textbf{Maximization to Find the Watermark-removed Counterparts.} Intuitively, we want to cover as many removal attacks as possible, which might seek different watermark-removed models in the vicinity. Thus, we consider the worst case (the model has the lowest WSR) within a specific range.
Given a feasible perturbation region $\mathcal{B} \triangleq \{\bm{\delta} \mid \Vert \bm{\delta}\Vert_2 \leq \epsilon \Vert\bm{\theta}\Vert_2 \}$, where $\epsilon >0 $ is a given perturbation budget, we attempt to find an adversarial parametric perturbation $\bm{\delta}$, 
\begin{equation}
    \bm{\delta} \leftarrow \max_{\bm{\delta} \in \mathcal{B}} \mathcal{L}(\bm{\theta} + \bm{\delta}, \mathcal{D}_w).
\end{equation}
In general, $\bm{\delta}$ is the worst-case weight perturbation that can be added to the watermarked model for generating its perturbed version $f_{\bm{\theta} + \bm{\delta}}(\cdot)$ with low watermark success rate.

\vspace{0.5em}
\noindent 
\textbf{Minimization to Recover the Watermark Behaviors.} After seeking the worst case in the vicinity, we should reduce the training loss on watermark samples of the perturbed model $f_{\bm{\theta} + \bm{\delta}}(\cdot)$ to recover its watermark behavior. Meanwhile, we always expect the model $f_{\bm{\theta}}(\cdot)$ to have low training loss on the clean training data to have satisfactory utility. Therefore, the training with watermark embedding is formulated as follows:
\begin{equation}
    \min_{\bm{\theta}} \big[ \mathcal{L}(\bm{\theta}, \mathcal{D}_c) + \alpha  \cdot \max_{\bm{\delta} \in \mathcal{B}} \mathcal{L}(\bm{\theta} + \bm{\delta}, \mathcal{D}_w) \big].
\end{equation}

\vspace{0.5em}
\noindent 
\textbf{The Perturbation Generation.} However, since DNN is severely non-convex, it is impossible to solve the maximization problem accurately. Here, we apply a single-step method to approximate the worst-case perturbation. Besides, the perturbation magnitude varies across architectures. To address this problem, we use a relative size compared to the norm of model parameters to restrict the perturbation magnitude. In conclusion, our proposed method to calculate the parametric perturbation is as follows: 
\begin{equation}
    % \bm{\delta} \leftarrow \epsilon \frac{\nabla_{\bm{\theta}} \mathcal{L}(\bm{\theta}, \mathcal{D}_w)}{\Vert \nabla_{\bm{\theta}} \mathcal{L}(\bm{\theta}, \mathcal{D}_w) \Vert} \Vert \bm{\theta} \Vert,
    \bm{\delta} \leftarrow \epsilon  \Vert \bm{\theta} \Vert_2 \cdot \frac{\nabla_{\bm{\theta}}  \mathcal{L}(\bm{\theta}, \mathcal{D}_w)}{\Vert \nabla_{\bm{\theta}} \mathcal{L}(\bm{\theta}, \mathcal{D}_w) \Vert}_2 ,
\end{equation}
where $\frac{\nabla_{\bm{\theta}}  \mathcal{L}(\bm{\theta}, \mathcal{D}_w)}{\Vert \nabla_{\bm{\theta}} \mathcal{L}(\bm{\theta}, \mathcal{D}_w) \Vert}_2$ is the normalized direction vector whose length equals 1, and $\epsilon  \Vert \bm{\theta} \Vert_2$ controls the magnitude of the perturbation in a relative way.

%------------------------------------------------------------------------
\subsection{Estimate BatchNorm Statistics on Clean Inputs}
\label{sec:domain_shift}

\noindent \textbf{The Assumption of Domain Shift.}
In preliminary experiments, we find our proposed algorithm cannot improve the robustness of the watermark (see Table \ref{tab:different-components}). We conjecture this failure is caused by the domain shift between the defense and attacks. Specifically, we only feed watermark samples into DNN, and all inputs of each layer are normalized by statistics from them when computing the adversarial perturbation and recovering the watermark behavior. In other words, the defender conducts the watermark embedding in the domain of watermark samples. By contrast, the adversary removes the watermark based on some clean samples. A similar problem about domain shift is also observed in domain adaption~\cite{li2016revisiting}.

\vspace{0.3em}
\noindent \textbf{The Verification of Domain Shift.} To verify the aforementioned assumption, we analyze the estimated mean and variance inside BatchNorm for clean samples and watermark samples. We visualize these estimations of different channels in the 9-th layer of ResNet-18~\cite{he2016deep} on CIFAR-10~\cite{krizhevsky2009learning}, and set the images with ``TEST'' as the watermark samples for the discussion. 
As shown in Figure~\ref{fig:disentangled-distribution}, there is a significant discrepancy between clean samples (the blue bar) and watermark samples (the orange bar). Since vanilla APP is performed using watermark samples while the attacker removes the watermark using clean samples, the discrepancy between clean and watermark samples may hinder the robustness of the watermark behavior.

\begin{figure}[!t]%靠文字内容的右侧
\centering
\vspace*{-0.5em}
\includegraphics[height=4.5cm]{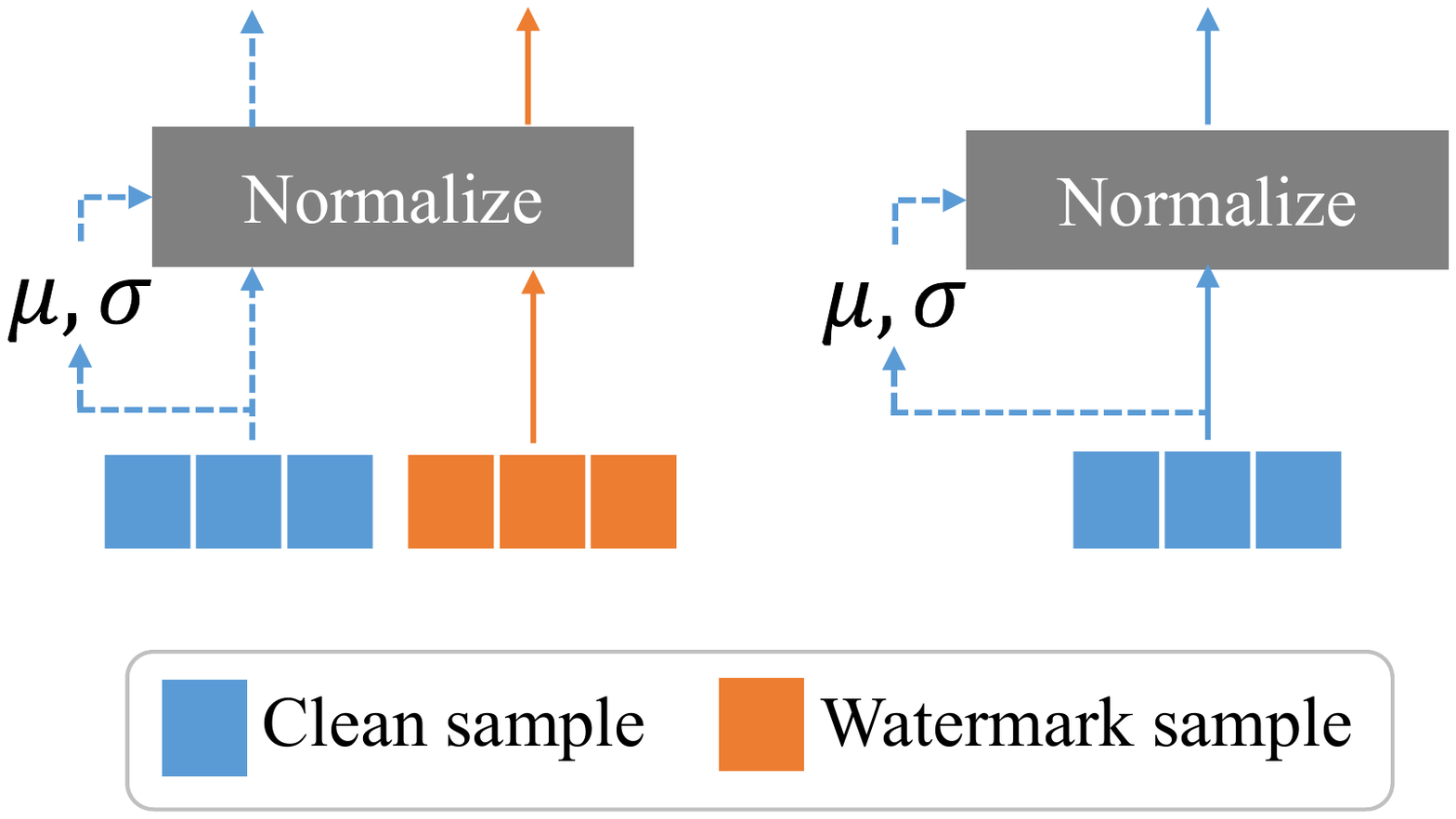}
\vspace{-0.8em}
    \caption{The diagram of our c-BN. We use BatchNorm statistics from the clean inputs to normalize the watermark inputs.}
  % \vspace{-1.5em}
  \label{fig:cbn}
\end{figure}

\vspace{0.3em}
\noindent \textbf{{The Proposed Customized BatchNorm}.} To reduce the discrepancy, we propose \textbf{c}lean-sample-based \textbf{B}atch\textbf{N}orm (c-BN). During forward propagation, we use BatchNorm statistics calculated from an extra batch of clean samples to normalize the watermark samples (the left part of Figure Figure~\ref{fig:cbn}), while we keep the BatchNorm unchanged for clean samples (the right part of Figure~\ref{fig:cbn}). In the implementation, since we always have a batch of clean samples $\mathcal{B}_c$ and a batch of watermark samples $\mathcal{B}_w$ for each update of model parameters, we always calculate the BatchNorm statistics and normalize inputs for each layer based on the clean batch $\mathcal{B}_c$. Thus, our APP-based watermarking training with c-BN can be reformulated as follows: 
\begin{equation}
    \min_{\bm{\theta}} \big[ \mathcal{L}(\bm{\theta}, \mathcal{D}_c) + \alpha  \cdot \max_{\bm{\delta} \in \mathcal{B}} \mathcal{L}(\bm{\theta}+\bm{\delta}, \mathcal{D}_w; \mathcal{D}_c)) \big],
\end{equation}
where $\mathcal{L}(\cdot, \cdot; \mathcal{D}_c)$ denotes that, when calculating this loss term, we use clean samples to estimate batch statistics during forward propagation in c-BN.

\subsection{The Overall Algorithm}

Here, we introduce the final algorithm of our method, which consists of adversarial parametric perturbation (APP) and clean-sample-based BatchNorm (c-BN). The pseudo-code of our method can be found in Algorithm~\ref{alg:embed-app}. Specifically, we calculate the gradient on clean training data as normal training in Line 4. In Line 6, we calculate the APP using clean batch statistics estimated by c-BN. Based on the APP, we calculate the gradient of the perturbed model on the watermarked data and add it to the gradient from clean data in Line 7. We update the model parameters in Line 8, and repeat the above steps until training converges.

\begin{figure*}[!htb]
\vspace{-1em}
	\centering
	\subfloat[\small{Original}]{\includegraphics[width=0.16\linewidth]{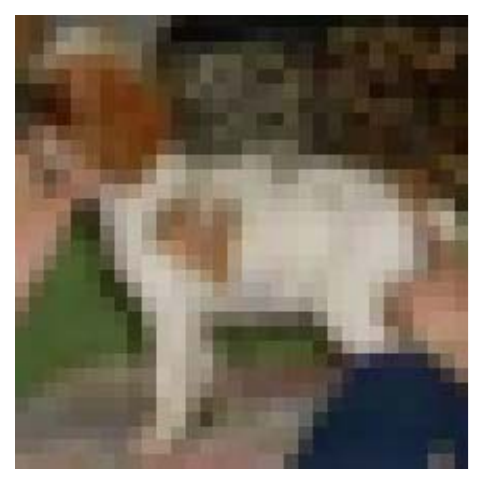}} \hspace{2.5em}
	\subfloat[\small{Content}]{\includegraphics[width=0.16\linewidth]{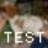}} \hspace{2.5em}
	\subfloat[\small{Noise}]{\includegraphics[width=0.16\linewidth]{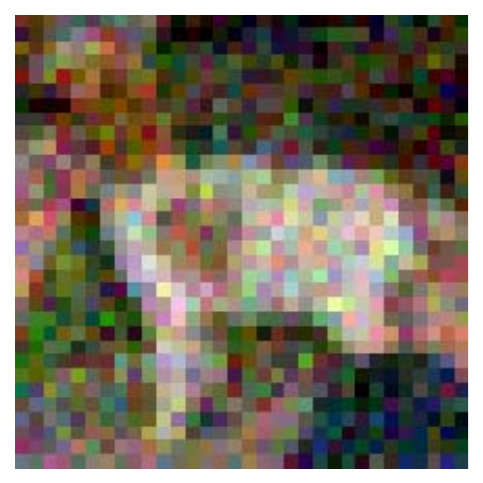}} \hspace{2.5em}
	\subfloat[\small{Unrelated}]{\includegraphics[width=0.16\linewidth]{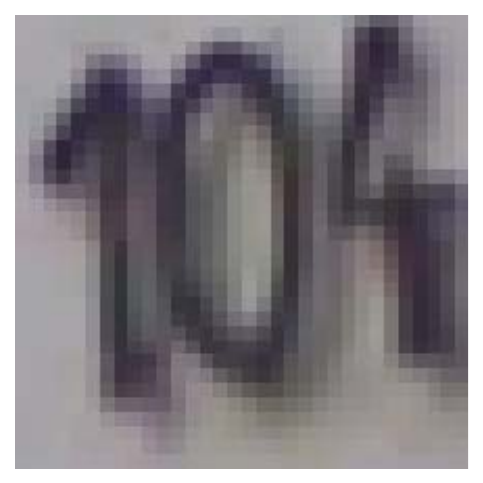}}
\vspace{0.5em}
\caption{The example of different watermark samples.}
	\label{fig:watermark-samples}
    % \vspace{-0.5em}
\end{figure*}

\begin{algorithm}[!t]
\renewcommand{\algorithmicrequire}{\textbf{Input:}}
	\renewcommand{\algorithmicensure}{\textbf{Output:}}
	\caption{Training APP-based watermarked model.}
	\label{alg:embed-app}
	\begin{algorithmic}[1]
		\REQUIRE Network $f_{\bm{\theta}}(\cdot)$, clean training set $\mathcal{D}_{c}$, watermarked training set $\mathcal{D}_{w}$, batch size $n$ for clean data, batch size $m$ watermarked data, learning rate $\eta$, perturbation magnitude $\epsilon$ 
		\STATE Initialize model parameters $\bm{\theta}$
            \REPEAT
                    \STATE Sample mini-batch $\mathcal{B}_{c}=\{(\bm{x}_1, y_1), \cdots, (\bm{x}_n, y_n)\}$ from $\mathcal{D}_{c}$
                    \STATE $\bm{g} \leftarrow \nabla_{\bm{\theta}} \mathcal{L}(\bm{\theta}, \mathcal{B}_c)$
		    \STATE Sample mini-batch $\mathcal{B}_{w}=\{(\bm{x}_1^\prime, y_1^\prime), \cdots, (\bm{x}_m^\prime, y_n^\prime)\}$ from $\mathcal{D}_{w}$
		    % \STATE $\bm{\delta} \leftarrow \epsilon \frac{\nabla_{\bm{\theta}} \mathcal{L}(\bm{\theta}, \mathcal{B}_w)}{\Vert \nabla_{\bm{\theta}} \mathcal{L}(\bm{\theta}, \mathcal{B}_w) \Vert} \Vert \bm{\theta} \Vert$
            \STATE $\bm{\delta} \leftarrow \epsilon \Vert \bm{\theta} \Vert_2 \frac{ \nabla_{\bm{\theta}} \mathcal{L}(\bm{\theta}, \mathcal{B}_w; \mathcal{B}_c)}{\Vert \nabla_{\bm{\theta}} \mathcal{L}(\bm{\theta}, \mathcal{B}_w; \mathcal{B}_c) \Vert} $ 
            % \STATE $\bm{g} \leftarrow \bm{g} + \nabla_{\bm{\theta}} [ \alpha \mathcal{L}(\bm{\theta} + \bm{\delta}, \mathcal{B}_w) ]$
            \STATE $\bm{g} \leftarrow \bm{g} + \nabla_{\bm{\theta}} [ \alpha \mathcal{L}(\bm{\theta} + \bm{\delta}, \mathcal{B}_w;\mathcal{B}_c)]${\quad// $\mathcal{L}(\cdot, \cdot; \mathcal{D}_c)$ denotes that, clean samples are used to estimate batch statistics during forward propagation in c-BN.}
		  \STATE $\bm{\theta} \leftarrow \bm{\theta} - \eta \bm{g}$  
            \UNTIL training converged
        \ENSURE Watermarked network $f_{\bm{\theta}}(\cdot)$
	\end{algorithmic}
	\label{alg:APP}
\end{algorithm}

%------------------------------------------------------------------------

\section{Experiments}

In this section, we conduct comprehensive experiments to evaluate the effectiveness of our proposed method, including a comparison with other watermark embedding schemes, ablation studies, and some exploratory experiments to understand our proposed method.

\subsection{Experiment Settings}

\noindent \textbf{Dataset Preparation.} We conduct experiments on CIFAR-10 and CIFAR-100~\cite{krizhevsky2009learning}.
To verify the effectiveness on more practical scenarios, we also do experiments on a subset of the ImageNet~\cite{deng2009imagenet} dataset, containing 100 classes with 50,000 images for training (500 images per class) and 5,000 images for testing (50 images per class). 
Similar to Zhang \etal~\cite{zhang2018protecting}, we consider three types of watermark samples: 1) Content: adding extra meaningful content to normal images (``TEST'' in our experiments). 2) Noise: adding a meaningless randomly-generated noise into normal images; 3) Unrelated: using images from an unrelated domain (SVHN~\cite{netzer2011reading} in our experiments). Figure \ref{fig:watermark-samples} visualizes samples for different watermark types. 
We set `0' as the target label, \textit{i.e.}, the watermarked DNN always predicts watermark samples as class ``airplane'' on CFIAR-10 and as ``beaver'' on CIFAR-100. We use 80\% of the original training data to train the watermarked DNNs and use the remaining 20\% for potential watermark-removal attacks. Before training, we replace 1\% of the current training data as the watermark samples.

\vspace{0.5em}
\noindent \textbf{Settings for Watermarked DNNs.} 
We train a ResNet-18 \cite{he2016deep} for 100 epochs with an initial learning rate of 0.1 and weight decay of $5\times10^{-4}$. The learning rate is multiplied by 0.1 at the 50-th and 75-th epoch. 
To train watermarked DNNs, we use our method and several state-of-the-art baselines: 1) \textit{vanilla} watermarking training~\cite{zhang2018protecting}; 2) exponentialized weight (EW) method~\cite{namba2019robust}; 3) the empirical verification\footnote{There is also a certified verification in ~\cite{bansal2022certified}, which requires full access to the parameters of the suspicious model. It is out of our scope and we only consider its empirical verification via API.} from certified watermarking (CW) ~\cite{bansal2022certified}. 
For our APP, we set the coefficient for watermark loss $\alpha=0.01$ and the maximum perturbation size $\epsilon=0.02$ on CIFAR-10 and CIFAR-100, and $\epsilon=0.01$ on ImageNet. Unless otherwise specified, we always use our c-BN during training.

\vspace{0.5em}
\noindent \textbf{Settings for Removal Attacks.} We evaluate the robustness of the watermarked DNN against several state-of-the-art watermark-removal attacks, including: 1) fine-tuning (FT)~\cite{uchida2017embedding}; 2) fine-pruning (FP)~\cite{liu2018fine}; 3) adversarial neural pruning (ANP)~\cite{wu2021adversarial}; 4) neural attention distillation (NAD)~\cite{li2021neural}; 5) mode connectivity repair (MCR)~\cite{zhao2020bridging}; 6) neural network laundering (NNL)~\cite{aiken2021neural}. 
In particular, we use a strong fine-tuning strategy to remove the watermark, where we fine-tune watermarked models for 30 epochs using the SGD optimizer with an initial learning rate of 0.05 and a momentum of 0.9. The learning rate is multiplied by 0.5 every 5 epochs. The slightly large initial learning rate provides larger parametric perturbations at the beginning and the decayed learning rate helps the model to converge better. More details about FT and other removal methods can be found in Appendix~\ref{app:watermark-removal}.

\begin{table*}[!t]
  \centering
%\small
  \caption{Performance (average over 3 random runs) of 3 watermark-injection methods and 3 types of watermark inputs against 6 removal attacks on CIFAR-10. \textit{Before}: BA/WSR of the trained watermarked models; \textit{After}: the remaining WSR after watermark-removal attacks. \textit{AvgDrop} indicates the average changes in WSR against all attacks.}
    \scalebox{0.93}{
    \begin{tabular}{ccccccccccc}
    
    \toprule
    \multirow{2}[2]{*}{Type} & \multirow{2}[2]{*}{Method} & \multicolumn{2}{c}{Before} & \multicolumn{6}{c}{After}                    & \multirow{2}[2]{*}{AvgDrop} \\
\cmidrule(lr){3-4} \cmidrule(lr){5-10}  &       & BA    & WSR   & FT    & FP    & ANP   & NAD   & MCR   & NNL   &  \\
    \midrule
    \multirow{4}[2]{*}{Content} & \textit{Vanilla}    & \textbf{93.86} & 99.56 & 56.78 & 74.58 & 25.34 & 48.14 & 16.56 & 21.02 & $\downarrow$ 59.15  \\
          & EW    & 92.86 & 99.17 & 55.11 & 63.22 & 66.24 & 48.92 & 25.17 & 29.15 & $\downarrow$ 51.20 \\
          & CW    & 93.73 & 99.62 & 26.98 & 54.22 & 27.39 & 29.18 & 29.97 & 19.78 & $\downarrow$ 68.36 \\
          & Ours  & 93.42 & \textbf{99.87} & \textbf{96.63} & \textbf{98.44} & \textbf{99.56} & \textbf{90.76} & \textbf{84.65} & \textbf{68.58} & $\downarrow$ \textbf{10.10} \\
    \midrule
    \multirow{4}[2]{*}{Noise} & \textit{Vanilla}    & 93.57 & 99.99 & 28.38 & 28.21 & 14.52 & 3.88  & 10.99 & 1.00  & $\downarrow$ 85.50 \\
          & EW    & 92.99 & 99.99 & 5.10  & 39.35 & 28.54 & 0.04  & 0.07  & \textbf{3.34} & $\downarrow$ 87.25 \\
          & CW    & \textbf{93.67} & \textbf{100.00} & 0.13  & 10.87 & 0.18  & 0.04  & 1.41  & 0.30 & $\downarrow$ 97.84\\
          & Ours  & 93.47 & \textbf{100.00} & \textbf{66.54} & \textbf{75.59} & \textbf{83.73} & \textbf{23.98} & \textbf{68.86} & 3.22  & $\downarrow$ \textbf{46.35} \\
    \midrule
    \multirow{4}[2]{*}{Unrelated} & \textit{Vanilla}    & \textbf{93.52} & \textbf{100.00} & 18.82 & 24.61 & 22.31 & 2.76  & 10.91 & 67.35 & $\downarrow$ 75.54 \\
          & EW    & 93.02 & 99.97 & 71.46 & 66.59 & 46.48 & 12.48 & 32.44 & 64.94 & $\downarrow$ 50.90 \\
          & CW    & 93.47 & \textbf{100.00} & 9.51  & 14.17 & 3.20  & 5.28  & 5.02  & 13.41 & $\downarrow$ 91.57 \\
          & Ours  & 93.30 & 99.95 & \textbf{96.15} & \textbf{95.46} & \textbf{99.60} & \textbf{89.28} & \textbf{87.49} & \textbf{94.49} & $\downarrow$ \textbf{6.20} \\
    \bottomrule
    \end{tabular}%
    }
  \label{tab:cifar-10}%
\end{table*}%

% Table generated by Excel2LaTeX from sheet 'main'
\begin{table*}[htbp]
    \centering
    %\small
  \caption{Performance (average over 3 random runs) of 3 watermark-injection methods and 3 types of watermark inputs against 6 removal attacks on ImageNet-subset. \textit{Before}: BA/WSR of the trained watermarked models; \textit{After}: the remaining WSR after watermark-removal attacks. \textit{AvgDrop} indicates the average changes in WSR against all attacks.}
    \scalebox{0.93}{
    \begin{tabular}{ccccccccccc}
    \toprule
    \multirow{2}[2]{*}{Type} & \multirow{2}[2]{*}{Method} & \multicolumn{2}{c}{Before} & \multicolumn{6}{c}{After}                    & \multirow{2}[2]{*}{AvgDrop} \\
\cmidrule(lr){3-4} \cmidrule(lr){5-10}  &       & BA    & WSR   & FT    & FP    & ANP   & NAD   & MCR   & NNL   &  \\
    \midrule
    \multirow{4}[1]{*}{Content} & \textit{Vanilla} & 74.81 & 98.26 & 22.18 & 9.31  & 43.91 & 4.40  & 12.48 & 28.05 & $\downarrow$ 78.20 \\
          & EW    & \textbf{75.15} & 95.85 & 8.95  & 3.82  & 17.07 & 3.02  & 8.82  & 19.96 & $\downarrow$ 85.58 \\
          & CW    & 74.52 & 99.05 & 6.35  & 0.16  & 0.26  & 0.68  & 2.92  & 17.91 & $\downarrow$ 94.34 \\
          & Ours  & 72.29 & \textbf{99.54} & \textbf{57.56} & \textbf{21.46} & \textbf{98.57} & \textbf{31.95} & \textbf{71.93} & \textbf{79.39} & $\downarrow$ \textbf{39.40} \\
    \midrule
    \multirow{4}[2]{*}{Noise} & \textit{Vanilla} & 74.47 & 98.65 & 9.54  & 2.79  & 29.00 & 9.75  & 8.06  & \textbf{3.60} & $\downarrow$ 88.20 \\
          & EW    & \textbf{75.09} & 95.36 & 3.58  & 4.08  & 1.19  & 1.62  & 4.19  & 1.56  & $\downarrow$ 92.66 \\
          & CW    & 74.11 & 98.32 & 15.35 & 2.57  & 11.65 & 5.65  & 3.41  & 2.56  & $\downarrow$ 91.45 \\
          & Ours  & 71.48 & \textbf{99.38} & \textbf{33.80} & \textbf{11.69} & \textbf{95.52} & \textbf{32.54} & \textbf{28.40} & 1.43  & $\downarrow$ \textbf{65.48} \\
    \midrule
    \multirow{4}[2]{*}{Unrelated} & \textit{Vanilla} & 74.69 & 99.97 & 47.40 & 36.53 & 99.66 & 24.16 & 54.43 & 30.87 & $\downarrow$ 51.13 \\
          & EW    & \textbf{75.25} & 99.97 & 33.64 & 31.12 & 94.40 & 59.91 & 12.94 & 56.70 & $\downarrow$ 51.85 \\
          & CW    & 74.97 & 99.99 & 38.94 & 0.86  & 1.97  & 43.68 & 65.74 & 26.66 & $\downarrow$ 70.34 \\
          & Ours  & 73.55 & \textbf{100.00} & \textbf{93.98} & \textbf{81.97} & \textbf{99.99} & \textbf{88.99} & \textbf{93.97} & \textbf{96.57} & $\downarrow$ \textbf{7.42} \\
    \bottomrule

    \end{tabular}%
    }
  \label{tab:imn-100}%
  %\vspace{-1em}
\end{table*}%

\vspace{0.5em}
\noindent \textbf{Evaluation Metrics.} We report the performance mainly on two metrics: \textbf{1)} watermark success rate (WSR) on watermark samples, that is the ratio of watermark samples that are classified as the target label by the watermarked DNN and \textbf{2)} benign accuracy (BA) on clean test data. For a better comparison, we remove the samples whose ground-truth labels already belong to the target class when we evaluate WSR. In general, an ideal watermark embedding method produces a model with high WSR and high BA, and keeps the high WSR after watermark-removal attacks.

%------------------------------------------------------------------------
\subsection{Main Results}\label{subsec:main-results}

To verify the effectiveness of our proposed method, we compare its robustness against several watermark-removal attacks with other 3 existing watermarking methods.  
All experiments are repeated over 3 runs with different random seeds. Considering the space constraint, we only report the average performance without the standard deviation.

\begin{table*}[!htb]
  \centering
  % \vspace{-1em}
  %\small
  \caption{The effect of the two components in our method.}
  % \vspace{-0.7em}
  % \footnotesize
    \scalebox{0.93}{
    \begin{tabular}{ccccccccccc} 
    \toprule
    
    \multirow{2}[2]{*}{APP} & \multirow{2}[2]{*}{c-BN} & \multicolumn{2}{c}{Before} & \multicolumn{6}{c}{After}                    & \multirow{2}[2]{*}{AvgDrop} \\
\cmidrule(lr){3-4} \cmidrule(lr){5-10}          &       & BA    & WSR   & FT    & FP    & ANP   & NAD   & MCR   & NNL   &  \\
    \midrule
     &      & 93.86 & 99.56 & 56.78 & 74.58 & 25.34 & 48.14 & 16.56 & 21.02 & $\downarrow$ 59.15 \\
         & $\checkmark$     & \textbf{93.94} & 99.75 & 58.14 & 74.92 & 10.26 & 35.17 & 19.14 & 23.37 & $\downarrow$ 62.91 \\
    % $\checkmark$     &      & 93.60 & 99.72 & 87.11 & 82.81 & 99.04 & 88.64 & 83.31 & 39.17 & $\downarrow$ 19.70 \\
    $\checkmark$     &      & 93.31 & 99.69 & 24.20 & 38.16 & 0.91 & 14.16 & 19.23 & 8.03 & $\downarrow$ 82.24 \\
    $\checkmark$     & $\checkmark$     & 93.42 & \textbf{99.87} & \textbf{96.63} & \textbf{98.44} & \textbf{99.56} & \textbf{90.76} & \textbf{84.65} & \textbf{68.58} & $\downarrow$ \textbf{10.10} \\
 
    \bottomrule
    \end{tabular}%
    }
  \label{tab:different-components}%
  % \vspace{-0.5em}
  
\end{table*}%

As shown in Table \ref{tab:cifar-10}, our APP-based method successfully embeds watermark behavior inside DNNs, achieving almost 100\% WSR with a negligible BA drop ($< 0.50 \%$). Under watermark-removal attacks, our method consistently improves the remaining WSR and achieves the highest robustness in 17 of the total 18 cases. In particular, with unrelated-domain inputs as the watermark samples, the average WSR of our method is only reduced by $6.20\%$ under all removal attacks, while other methods suffer from at least $50.90\%$ drop in WSR. We find that, although NNL is the strongest removal attack (all WSRs decrease below $22\%$) when watermark samples are those images superimposed by some content or noise, it has an unsatisfactory performance to unrelated-domain inputs as watermark samples\footnote{This is because NNL first reconstructs the watermark trigger (\textit{e.g.}, the content ``TEST'' on watermark samples) and then removes watermark behaviors. By contrast, when we use unrelated-domain inputs as watermark samples, there is no trigger pattern, leading to the failure of NNL.}. Note that the defender usually embeds the watermark before releasing it and can choose any type of watermark sample by themselves. Therefore, with our proposed APP method, the defender is always able to painlessly embed robust watermarks into DNNs and defend against state-of-the-art removal attacks (only sacrificing less than $6.2\%$ of WSR after attacks). We have similar findings on ImageNet (see Table~\ref{tab:imn-100}) and CIFAR-100 (see Appendix~\ref{app:cifar-100}).

%------------------------------------------------------------------------
\subsection{Ablation Studies} \label{subsec:ablation-study}

In this section, we conduct several experiments to explore the effect of each part in our proposed methods, including different components, varying perturbation magnitudes, and various target classes.
In the following experiments, we always use the images with the content ``TEST'' as the watermark sample unless otherwise specified.

\begin{figure}[t]
\captionsetup{justification=raggedright}  
\vspace{-0.8em}
  \centering  
	\subfloat{\label{fig:varying-eps-na}\includegraphics[height=3.5cm]{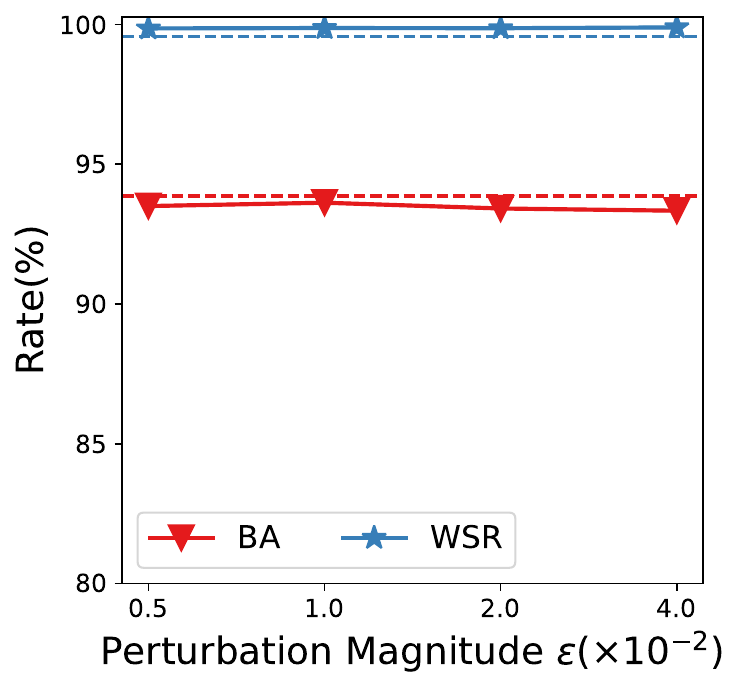}}
	\subfloat{\label{fig:varying-eps-atk}\includegraphics[height=3.5cm]{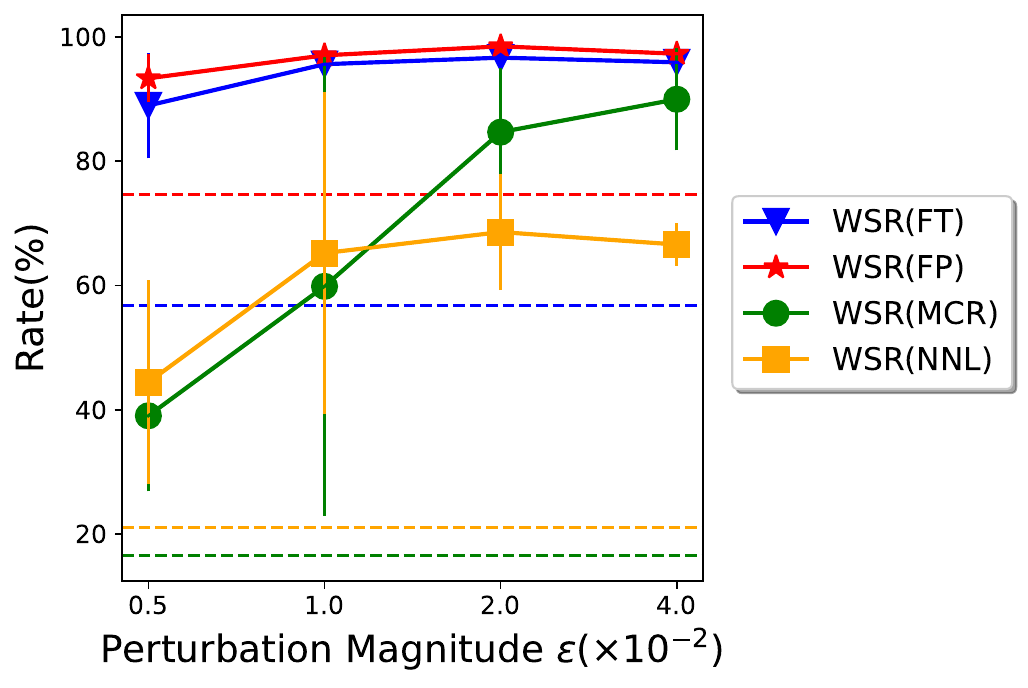}}
    \caption{The results with various magnitude $\epsilon$. We use the dashed line with the same color to show the performance when $\epsilon=0$. \textit{Left}: before attacks; \textit{Right}: after attacks.}
  \label{fig:varying-eps}
  \vspace{-0.5em}
\end{figure}

\vspace{0.3em}
\noindent \textbf{Effect of Different Components.} Our method consists of two parts, \textit{i.e.}, the adversarial parametric perturbation (APP) and the clean-sample-based BatchNorm (c-BN). we evaluate the contribution of each component. We train and evaluate watermarked DNNs without any components (the \textit{Vanilla} method), with one of the components, and with both components (our proposed method). In Table \ref{tab:different-components}, only with APP, we fail in keeping the average WSR under removal attacks due to the domain shift as mentioned in Sec \ref{sec:domain_shift}. Fortunately, with c-BatchNorm, APP solves the domain shift problem and successfully improves the robustness against removal attacks, \textit{e.g.}, it keeps WSR $>90\%$ against several removal attacks (FT, FP, ANP, and NAD), and even keeps WSR $68.58\%$ against the strongest attack NNL. Besides, we find the watermark with only c-BN fails to improve the WSR, which indicates the c-BN just helps APP rather than improving watermark robustness directly. In conclusion, both are essential components contributing to final robustness against watermark-removal attacks.

\begin{figure}[t]
\captionsetup{justification=raggedright}  
  \centering  
  \centering  
  \includegraphics[height=4cm]{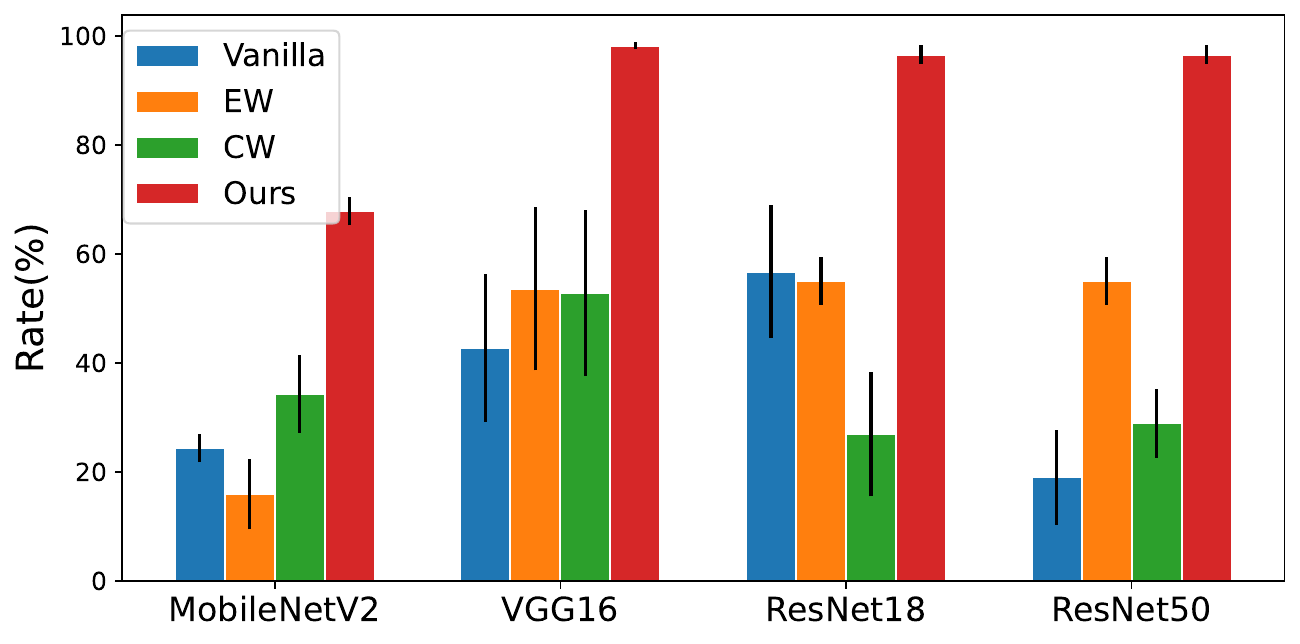}

  \caption{The results of our methods and other baselines with various architectures against FT attack. Our method consistently improves watermark robustness.}
  \label{fig:ft-varying-arch}
  \vspace{-1.2em}
\end{figure}

\begin{figure*}[!htb]
\vspace{-0.3cm}
	\centering
	\subfloat[Vanilla]{\label{fig:bd-tsne}\includegraphics[height=3.6cm]{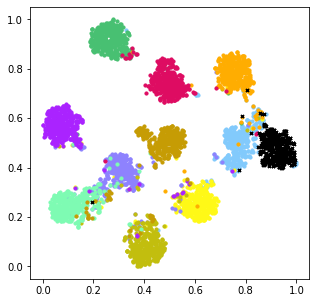}}
	\subfloat[EW]{\label{fig:ew-tsne}\includegraphics[height=3.6cm]{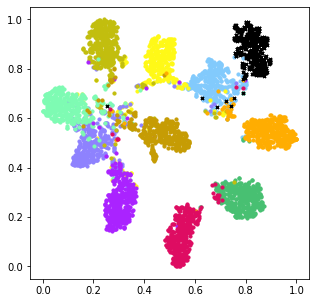}}
	\subfloat[CW]{\label{fig:cw-tsne}\includegraphics[height=3.6cm]{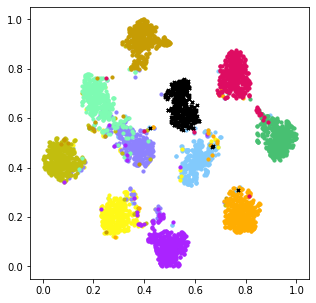}}
	\subfloat[Ours]{\label{fig:ours-tsne}\includegraphics[height=3.6cm]{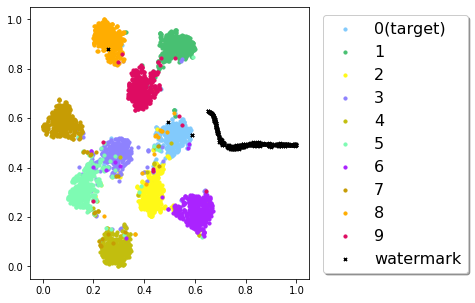}}
        \vspace{0.5em}
	\caption{The t-SNE visualization of hidden feature representations.}
        \vspace{-1em}
	\label{fig:compare-tsne}
\end{figure*}

\vspace{0.3em}
\noindent \textbf{Effect of Varying Perturbation Magnitude.} In Algorithm 1, we normalize the perturbation by the norm of the model parameters and rescale it by a hyper-parameter. Here, we explore the effect of this relative perturbation magnitude hyper-parameter $\epsilon$. We illustrate the performance of the watermarked DNNs before and after removal attacks with varying perturbation magnitude
in Figure \ref{fig:varying-eps}, and find that,  within a specific region $\epsilon \leq 4.0 \times 10^{-2}$, our method always improves the robustness against attacks while keeping BA high in a large range for hyperparameter. Besides, we find the selection of hyper-parameter $\epsilon$ is more related to the watermark embedding method itself rather than removal attacks (we have similar trends against FT, FP, MCR and NNL). This makes the selection of hyper-parameter $\epsilon$ quite straightforward and gives us simple guidance for tuning $\epsilon$ in practical scenarios:
Although knowing nothing about the potential attack (suppose the adversary applies MCR), the defender could tune the hyper-parameter against the FT attacks, and the resulting model also achieves satisfactory results against MCR. Detailed results against other attacks can be found in Appendix~\ref{app:varying-eps}.

\vspace{0.3em}
\noindent \textbf{Effect of Various Target Classes.} Recall that we have studied the effects of different watermark samples (Content, Noise, and Unrelated in Section~\ref{subsec:main-results}), here we further evaluate the effects of the different target classes as which the model classifies these watermark samples. We set the target class as 1, 2, 3, and 4, respectively. We obtain an average WSR of $94.87\%$, $79.81\%$, $84.36\%$ and $87.76\%$ respectively under all removal attacks, while the \textit{vanilla} method only achieves $32.91\%, 20.79\%, 32.28\%$, and $10.13\%$ (details can be found in Appendix~\ref{app:varying-target}). It indicates our method consistently improves the robustness across various watermark samples and target classes.

\begin{figure}[]
% 	\centering
    \centering
	\subfloat[\small{Size}]{\includegraphics[height=3.7cm]{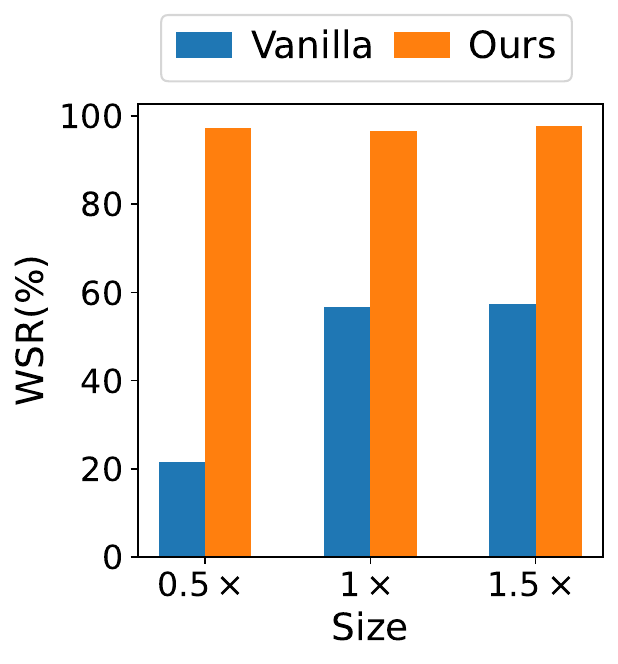}}
	\subfloat[\small{Transparency}]{\includegraphics[height=3.7cm]{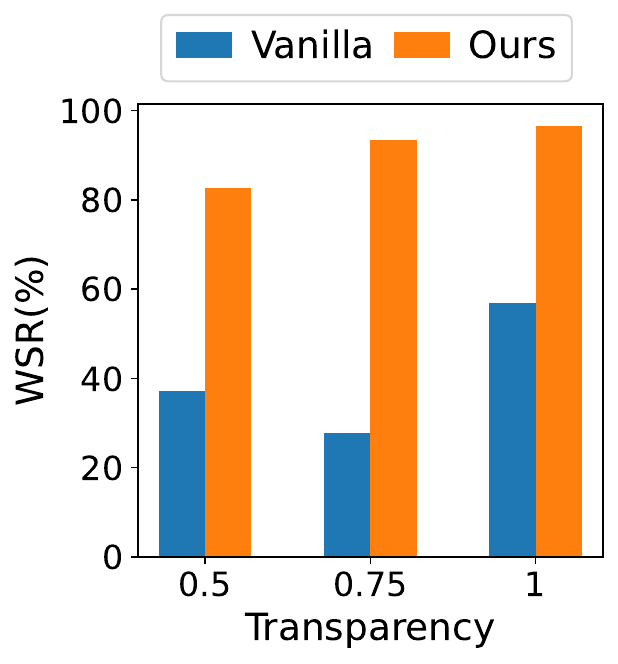}}
    \vspace{0.5em}
	\caption{Results with various trigger sizes and transparencies. $1\times$ represents the settings of the original trigger. }
    \vspace{-1em}
	\label{fig:font-transparency}
\end{figure}

\vspace{0.3em}
\noindent \textbf{Effect on Trigger Size and Transparency.} 
To further verify that our method can apply to triggers with different sizes and transparencies, we also exploit various sizes and transparencies of the “TEST” trigger and evaluate the robustness using FT attack. As shown in Figure~\ref{fig:font-transparency}, our method consistently reaches better performance than the baseline across various trigger sizes and transparencies.

\vspace{0.3em}
\noindent \textbf{Effect of Different Architectures.} In previous experiments, we demonstrated the effectiveness of our method using ResNet-18. Here, we explore the effect of the model architectures across different sizes including MobileNetV2 
~\cite{sandler2018mobilenetv2} (a tiny model), VGG16~\cite{simonyan2014very}, ResNet-18 and ResNet-50~\cite{he2016deep} (a relatively large model) with same hyper-parameters (especially $\epsilon$). As shown in Figure~\ref{fig:ft-varying-arch}, our method always achieves notable improvements ($> 30\%$) compared with other baseline methods in all cases.

%------------------------------------------------------------------------

\subsection{A Closer Look at the APP Method}
In this section, we further explore the mechanism of our APP. We visualize the landscape of watermarked model in the parameter space and the distribution of the clean and watermark samples in the feature space for discussions.

\begin{figure}[!tb]
	\centering
 %\vspace{-0.8cm}
 % \hspace{-0.5cm}
    \subfloat[\small{Watermark Success Rate}]{\includegraphics[height=3.7cm]{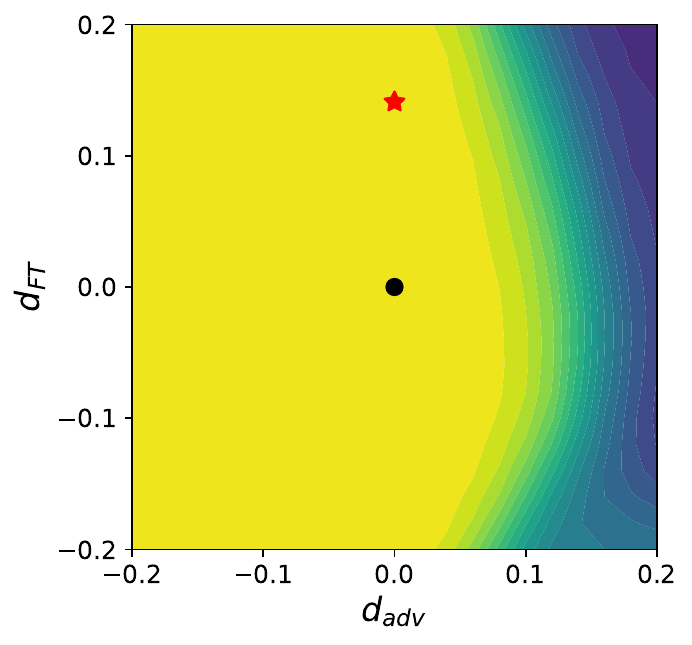}}
	\subfloat[\small{Benign Accuracy}]{\includegraphics[height=3.7cm]{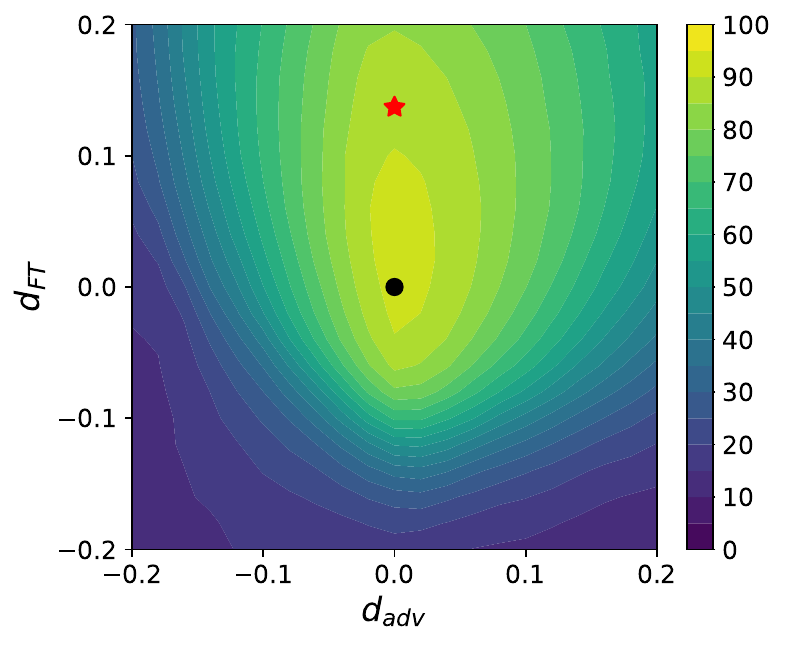}}
    \vspace{0.5em}
	\caption{The performance of models in the vicinity of APP-based watermarked model in the parameter space. $d_{FT}$ denotes the direction of fine-tuning and $d_{adv}$ denotes the adversarial direction. \textit{black dot}: the original watermarked model; \textit{red star}: the model after fine-tuning. }
	\label{fig:ours-vicinity}
    \vspace{-1em}
\end{figure}

\vspace{0.3em}
\noindent \textbf{The Parameter Space.} We start by studying the properties of the watermarked model in the parameter space in the Introduction and illustrate how WSR changes in the vicinity of the watermarked model from the \textit{vanilla} method in Figure \ref{fig:intro}. Here, we use the same visualization method to show the vicinity of the APP-based method (please see more details in Appendix~\ref{app:plot}). As shown in Figure \ref{fig:ours-vicinity}, we find the APP-based watermarked model is able to keep WSR high within a larger range compared to the \textit{vanilla} one. Especially, our model is better in robustness against parametric perturbation along the adversarial direction, which makes it more difficult for the adversary to find watermark-removed models in the vicinity of the protected model.

\vspace{0.3em}
\noindent \textbf{The Feature Space.} To dive into APP, we also visualize the hidden representation of clean samples and watermark samples using the t-SNE method~\cite{van2008visualizing} based on different watermark embedding schemes. As shown in Figure~\ref{fig:compare-tsne}, in the feature space of our model, the cluster of watermark samples in our method has a larger coverage in the feature space. This may explain why our method is more robust because moving all these watermark samples back to their original clusters takes much more effort. Implementation details and more results can be found in Appendix~\ref{app:t-SNE}.
%------------------------------------------------------------------------

\section{Discussion and Conclusion}

In our threat model, we actually limit the parameter perturbation size, \textit{i.e.}, the adversary cannot change the model parameters too much. By contrast, in practice, the adversary is only required to maintain the high benign accuracy of DNNs during watermark-removal attacks. We admit the latter is a better threat model, while it is infeasible to analyze rigorously. It is mostly because we cannot explicitly describe the relationship between benign accuracy and model parameters (we only know some checkpoints and their BA), which prevents its direct usage in the algorithm. Instead, we use a simplified constraint by the perturbation magnitude and believe it is a feasible method: (1) In most cases, attackers use the watermarked model as the initial point and fine-tune model parameters, which (probably) bounds the change of model parameters within a distance; {(2) We achieve better robustness against various practical attacks using our threat model. We notice that the defense in our threat model is only a prerequisite for defense in the better threat model. We hope our method can serve as the cornerstone towards truly robust watermarks.

Overall, we investigated the parameter space of watermarked DNNs in this paper. We found that there exist many watermark-removed models in the vicinity of the watermarked model, which may be easily used by removal attacks. To alleviate this problem, we proposed a minimax formulation to find the watermark-removed models and repair their watermark behaviors. In particular, we observed that there is a domain shift between defenses and removal attacks when calculating BatchNorm statistics, based on which we proposed to estimate them only with benign samples (dubbed `c-BN'). We conducted extensive experiments on benchmark datasets, showing that our method can consistently improves the robustness against several state-of-the-art removal attacks. We hope our method could help model owners better protect their intellectual properties.%, to facilitate DNN sharing or trading.

\section*{Acknowledgement}

This work is supported in part by the National Natural Science Foundation of China under Grant 62171248, Shenzhen Science and Technology Program under Grant JCYJ20220818101012025, and the PCNL Key Project under Grant PCL2021A07.

%------------------------------------------------------------------------

% \clearpage
{\small
\bibliographystyle{ieee_fullname}
\bibliography{egbib}
}

%------------------------------------------------------------------------
%------------------------------------------------------------------------
\clearpage
\appendix
%------------------------------------------------------------------------
\section{Details about Vicinity Visualization} \label{app:plot}
To visualize the vicinity, we measure the watermark success rate (WSR) and benign accuracy (BA) on the panel spanned by the two directions $d_{adv}$ and $d_{FT}$. Specifically, $d_{adv}$ is the direction to erase watermark, \textit{i.e.}, $d_{adv}=\nabla_{\bm{\theta}} \mathcal{L}(\bm{\theta}, \mathcal{D}_w)$, and $d_{FT}$ is the direction from the original watermarked model $\bm{\theta}_w$ to a fine-tuned model $\bm{\theta}_{FT}$, \textit{i.e.}, $d_{FT}=\bm{\theta}_{FT}-\bm{\theta}_w$. We fine-tune the original model $\bm{\theta}_w$ for 40 iterations with the SGD optimizer using a learning rate 0.05 to obtain $\bm{\theta}_{FT}$. We explore the vicinity by moving the original parameter along with these two directions, recoding WSR and BA of neighbor model. For easier comparison, we use the relative distance in the parametric space, \textit{i.e.},
\begin{equation}
    \bm{\theta} = \bm{\theta}_w + \alpha \frac{d_{adv}}{\Vert d_{adv} \Vert}  \Vert \bm{\theta}_w  \Vert  + \beta \frac{d_{FT}}{\Vert d_{FT} \Vert}  \Vert \bm{\theta}_w \Vert,
\end{equation}
where $(\alpha,\beta)$ are the given coordinates. After obtaining the parameter $\bm{\theta}$ in the vicinity, we further adjust BatchNorm by re-calculating the statistic on the clean dataset to restore benign accuracy. Finally, we evaluate this neighbor model and record its benign accuracy and watermark success rate.

%------------------------------------------------------------------------
%------------------------------------------------------------------------
\section{Details about Main Experiments} \label{app:exp-details}
In this section, we first briefly introduce our baseline methods, then provide the detailed settings for our main experiments. We report the full results on CIFAR-10 and CIFAR-100 at the end. 
\subsection{More details about baseline methods} \label{app:baselines}
Vanilla model watermark~\cite{zhang2018protecting} mixed the watermark samples with the clean samples, based on which to train the model. EW~\cite{namba2019robust} trained the model with exponentially reweighted parameter $EW(\theta,T)$ rather than vanilla weight $\theta$. They exponentially reweighted the $i$th element of the $l$th parameter $\theta^l$, \textit{i.e.}, 
\begin{equation}
    EW(\theta^l,T) = \theta^l_{exp},\  \text{where}\ \ \theta^l_{exp,i} = \frac{\exp(|\theta_i^l|T)}{\max_i(\exp(|\theta_i^l|T))}\theta_i^l,
\end{equation}
and $T$ is a hyper-parameter adjusting the intensity of the reweighting. As shown in the above equation, the weight elements with a big absolute value will remain almost the same after the reweight operation, while the ones with a small value will decrease to nearly zero. This encourages the neural network to lean on the weights with large absolute values to make decisions, hence making the prediction less sensitive to small weight changes. CW~\cite{bansal2022certified} aimed at embedding a watermark with certifiable robustness. They adopted the theory of randomized smoothing~\cite{cohen2019certified} and watermarked the network using a gradient estimated with random perturbed weights. The gradient on the watermark batch $\mathcal{B}$ is calculated by 
\begin{equation}
    % E(x;y)2B[r l(x; y; + G)]
    g_\theta = \frac{1}{k}\sum_{i=1}^{k}E_{G\in\mathcal{N}(0,(\frac{i}{k})^2I)} E_{(x,y)\in\mathcal{B}} [\nabla l(x,y;\theta+G)],
\end{equation}
where $\sigma$ is the noise strength.
%------------------------------------------------------------------------
\subsection{Details about Watermark-removal Attacks}

Currently, there are some watermark-removal attacks to counter model watermarking. According to Lukas \etal~\cite{lukas2021sok}, existing removal attacks can be divided into three main categories, including 1) \emph{input pre-processing}, 2) \emph{model extraction}, and 3) \emph{model modification}. 
In general, the first type of attack pre-processes each input sample to remove trigger patterns before feeding it into the deployed model~\cite{lin2019invert}. 
Model extraction~\cite{hinton2015distilling,shafieinejad2021robustness} distills the dark knowledge from the victim model to remove distinctive prediction behaviors while preserving its main functionalities. 
Model modification~\cite{uchida2017embedding,liu2018fine} changes model weights while preserving its main structure. In this paper, we mainly focus on the model-modification-based removal attacks, since input pre-processing has minor benefits for countering backdoor-based watermark~\cite{lukas2021sok} and model extraction usually requires a large number of training samples that are inaccessible for defenders in practice~\cite{lukas2020deep}.

Apart from these traditional watermark attacks mentioned above, we also adopted some backdoor-removal methods to conduct a more thorough evaluation because our watermark method is backdoor-based. The backdoor-removal method can also be derived into two categories, including 1) \emph{post-training backdoor removal methods}~\cite{wu2020adversarial,li2021neural,zhao2020bridging} that remove backdoor with local benign samples after training, 2) \emph{training-time backdoor removal methods}~\cite{li2021anti,huang2021backdoor,gao2023backdoor} that directly train a clean model from a poisoned training set.  In our experiments, we focus on the \emph{post-training backdoor removal methods} because only the model owner controls the training process.

The description of our adopted watermark/backdoor-removal methods is listed in the following.

\vspace{0.3em}\noindent \textbf{FT.} Uchida \etal~\cite{uchida2017embedding} removed the watermark by updating model parameters using additional holding clean data.

\vspace{0.3em}\noindent \textbf{FP.} Liu \etal~\cite{liu2018fine} presumed that watermarked neurons are less activated by clean data, and thus pruned the least activated neurons in the last layer before fully-connected layers. They further find-tuned the pruned model to restore benign accuracy and suppress watermarked neurons.

\vspace{0.3em}\noindent \textbf{ANP.} Wu \etal~\cite{wu2021adversarial} found that backdoored neurons are sensitive to weight perturbation and proposed to prune these neurons to remove the backdoor.

\vspace{0.3em}\noindent \textbf{NAD.} Li \etal~\cite{li2021neural} utilized knowledge from a fine-tuned model where the watermark is partially removed, to guide the watermark unlearning.

\vspace{0.3em}\noindent \textbf{MCR.} Zhao \etal~\cite{zhao2020bridging} found that the existence of a high accuracy pathway connecting two backdoored models in the parametric space, and the interpolated model along the path usually doesn't have backdoors. This property allows MCR to be applied in the mission of watermark removal. 

\vspace{0.3em}\noindent \textbf{NNL.} Aiken \etal~\cite{aiken2021neural} first reconstructed trigger using Neural Cleanse~\cite{wang2019neural}, then reset neurons that behave differently on clean data and reconstructed trigger data, and further fine-tuned the model to restore benign accuracy and suppress watermarked neurons.
%------------------------------------------------------------------------
\subsection{More Details about Watermark Settings}\label{app:watermark-settings}
\vspace{0.3em}\noindent \textbf{Settings for EW.} As suggested in its paper~\cite{namba2019robust}, we fine-tune a pre-trained model to embed the watermark. We pre-train the model using the original dataset without injecting the watermark samples. The pre-trained model is trained for 100 epochs using the SGD optimizer with an initial learning rate of 0.1, the learning rate decays by a factor of 10 at the 50th and 75th epochs. We fine-tune the pre-trained model for 20 epochs to embed the watermark, with an initial learning rate of 0.1, and the learning rate is drop by 10 at the 10th and 15th epochs. 

\vspace{0.3em}\noindent \textbf{Settings for CW.} For a fair comparison, we adopt a learning rate schedule and a weight-decay factor identical to other methods. Unless otherwise specified, other settings are the same as those used in~\cite{bansal2022certified}.

\vspace{0.3em}\noindent \textbf{Settings for Our Method.}
For the classification loss term, we calculate the loss using a batch of 128 clean samples, while for the watermark term, we use a batch of 64 clean samples and 64 watermark samples to obtain the estimation of adversarial gradients.

\subsection{Details about Watermark-removal Settings}\label{app:watermark-removal}
\vspace{0.3em}\noindent \textbf{Settings for FT.} We fine-tune the watermarked model for 30 epochs using the SGD optimizer with an initial learning rate of 0.05 and a momentum of 0.9, the learning rate is dropped by a factor of 0.5 every five epochs.

\vspace{0.3em}\noindent \textbf{Settings for FP.} We prune 90\%  of the least activated neurons in the last layer before fully-connected layers, and after pruning, we fine-tune the pruned model using the same training scheme as FT. 

\vspace{0.3em}\noindent \textbf{Settings for ANP.} We set the pruning rate to 0.6, where all defense shares a similar BA, as shown in Figure~\ref{fig:anp-varying-threshold}. 

\vspace{0.3em}\noindent \textbf{Settings for NAD.} The original NAD only experimented on WideResNet models. In our work, we calculate the NAD loss over the output of the four main layers of ResNet, with all $\beta$s set to 1500. To obtain a better watermark removal performance, we use an initial learning rate of 0.02 , which is larger than 0.01 in the original paper~\cite{li2021neural}.

\vspace{0.3em}\noindent \textbf{Settings for MCR.} MCR finds a backdoor-erased model on the path connecting two backdoored models. But in our settings, only one watermarked model is available. Hence the attacker must obtain the other model via fine-tuning the original watermarked model, then perform MCR using the original watermarked model and fine-tuned model. We split the attacker's dataset into two equal halves, one used to fine-tune the model and the other one to train the curve connecting the original model and the fine-tuned model. This fine-tuning is performed for 50 epochs with an initial learning rate of 0.05, which decays by a factor of 0.1 every 10 epochs. For MCR results, $t=0$ denotes the original model and $t=1$ denotes the original model. We select results with $t=0.9$, where all defense shares similar BA, see Figure~\ref{fig:mcr-varying-threshold}. 

\vspace{0.3em}\noindent \textbf{Settings for NNL.} We reconstruct the trigger using Neural Cleanse~\cite{wang2019neural} for 15 epochs, and reset neurons that behave significantly different under clean input and reconstructed input, we fine-tune the model for 15 epochs with the SGD optimizer, the initial learning rate is 0.02 and is divided by 10 at the 10th epoch. 

%------------------------------------------------------------------------
\subsection{Detailed Results on CIFAR-10}
The detailed results on CIFAR-10 are shown in Table~\ref{tab:cifar-10-full}. Moreover, we can observe from Figure~\ref{fig:mcr-varying-threshold} and Figure~\ref{fig:anp-varying-threshold} that our method is more robust than other methods, regardless of the threshold value used in MCR and ANP. 

\begin{table*}[!htbp]
  \centering
  \small
  \caption{Results on CIFAR-10. `NA' denotes `No Attack'.}
  % \vspace{-0.7em}
  \scalebox{0.98}{
    \begin{tabular}{ccccccccccc}
    \toprule
    Metric & Type  & Method & NA    & FT    & FP    & ANP   & NAD   & MCR   & NNL & AvgDrop \\
    \midrule
    \multirow{12}[6]{*}{WSR} & \multirow{4}[2]{*}{Content} & Vanilla    & 99.56 & 56.78 & 74.58 & 25.34 & 48.14 & 16.56 & 21.02 & $\downarrow$59.15 \\
          &       & EW    & 99.17 & 55.11 & 63.22 & 66.24 & 48.92 & 25.17 & 29.15 & $\downarrow$51.20 \\
          &       & CW    & 99.62 & 26.98 & 54.22 & 27.39 & 29.18 & 29.97 & 19.78 & $\downarrow$68.36 \\
          &       & Ours  & \textbf{99.87} & \textbf{96.63} & \textbf{98.44} & \textbf{99.56} & \textbf{90.76} & \textbf{84.65} & \textbf{68.58} & $\downarrow$\textbf{10.10} \\
\cmidrule(lr){2-11}          & \multirow{4}[2]{*}{Noise} & Vanilla    & 99.99 & 28.38 & 28.21 & 14.52 & 3.88  & 10.99 & 1.00 & $\downarrow$ 85.50 \\
          &       & EW    & 99.99 & 5.10  & 39.35 & 28.54 & 0.04  & 0.07  & \textbf{3.34} & $\downarrow$ 87.25 \\
          &       & CW    & \textbf{100.00} & 0.13  & 10.87 & 0.18  & 0.04  & 1.41  & 0.30 & $\downarrow$ 97.84 \\
          &       & Ours  & \textbf{100.00} & \textbf{66.54} & \textbf{75.59} & \textbf{83.73} & \textbf{23.98} & \textbf{68.86} & 3.22 & $\downarrow$ \textbf{46.35} \\
\cmidrule(lr){2-11}           & \multirow{4}[2]{*}{Unrelated} & Vanilla    & \textbf{100.00} & 18.82 & 24.61 & 22.31 & 2.76  & 10.91 & 67.35 & $\downarrow$ 75.54 \\
          &       & EW    & 99.97 & 71.46 & 66.59 & 46.48 & 12.48 & 32.44 & 64.94 & $\downarrow$ 50.90 \\
          &       & CW    & \textbf{100.00} & 9.51  & 14.17 & 3.20  & 5.28  & 5.02  & 13.41 & $\downarrow$ 91.57 \\
          &       & Ours  & 99.95 & \textbf{96.15} & \textbf{95.46} & \textbf{99.60} & \textbf{89.28} & \textbf{87.49} & \textbf{94.49} & $\downarrow$ \textbf{6.20} \\
    \midrule
    \midrule
    \multirow{12}[6]{*}{BA} & \multirow{4}[2]{*}{Content} & Vanilla    & \textbf{93.86} & 91.80 & 92.19 & 90.15 & 90.39 & 89.27 & 91.92 & 2.91 \\
          &       & EW    & 92.86 & 90.95 & 91.45 & 89.41 & 88.72 & 88.31 & 91.14 & 2.87 \\
          &       & CW    & 93.73 & 91.75 & 91.99 & 89.67 & 90.29 & 89.00 & 91.77 & 2.98 \\
          &       & Ours  & 93.42 & 91.72 & 91.81 & 88.86 & 89.79 & 89.08 & 91.06 & 3.03 \\
\cmidrule(lr){2-11}           & \multirow{4}[2]{*}{Noise} & Vanilla    & 93.57 & 92.00 & 92.12 & 89.87 & 90.59 & 89.41 & 91.58 & 2.64 \\
          &       & EW    & 92.99 & 91.05 & 91.41 & 89.09 & 88.81 & 88.39 & 91.14 & 3.01\\
          &       & CW    & \textbf{93.67} & 91.19 & 91.79 & 86.32 & 85.12 & 88.74 & 91.28 & 4.60 \\
          &       & Ours  & 93.47 & 91.59 & 91.87 & 86.75 & 90.14 & 89.18 & 90.73 & 3.43\\
\cmidrule(lr){2-11}           & \multirow{4}[2]{*}{Unrelated} & Vanilla    & \textbf{93.52} & 91.53 & 91.91 & 90.16 & 89.16 & 88.22 & 90.77 & 3.23\\
          &       & EW    & 93.02 & 91.17 & 91.44 & 89.23 & 89.13 & 88.30 & 90.80 & 3.01\\
          &       & CW    & 93.47 & 91.17 & 91.29 & 86.31 & 88.97 & 87.83 & 90.72 & 4.60\\
          &       & Ours  & 93.30 & 91.47 & 91.46 & 86.48 & 89.70 & 89.08 & 90.36 & 3.54\\
    \bottomrule
    \end{tabular}%
    }
  \label{tab:cifar-10-full}%
\end{table*}%

\begin{figure*}[!htbp]
    \centering
	\subfloat[\small{Content}]{\includegraphics[height=3.35cm]{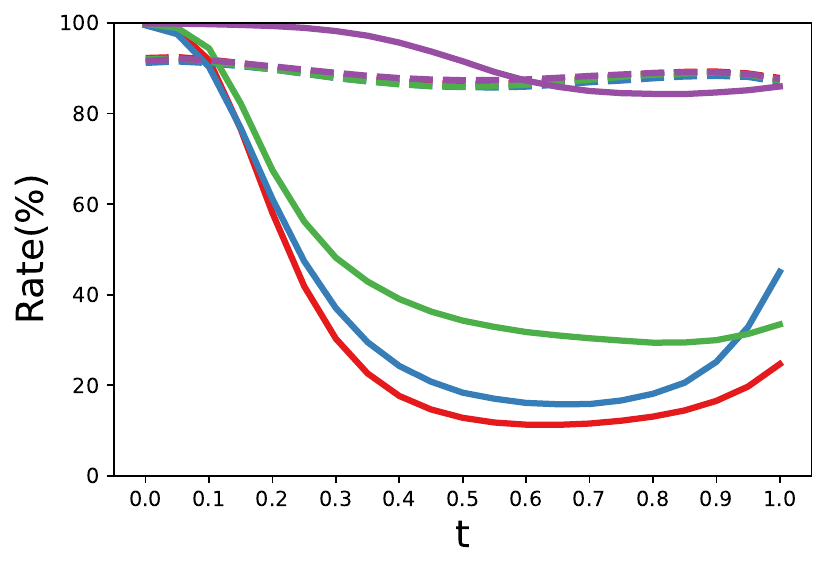}}
	\subfloat[\small{Noise}]{\includegraphics[height=3.35cm]{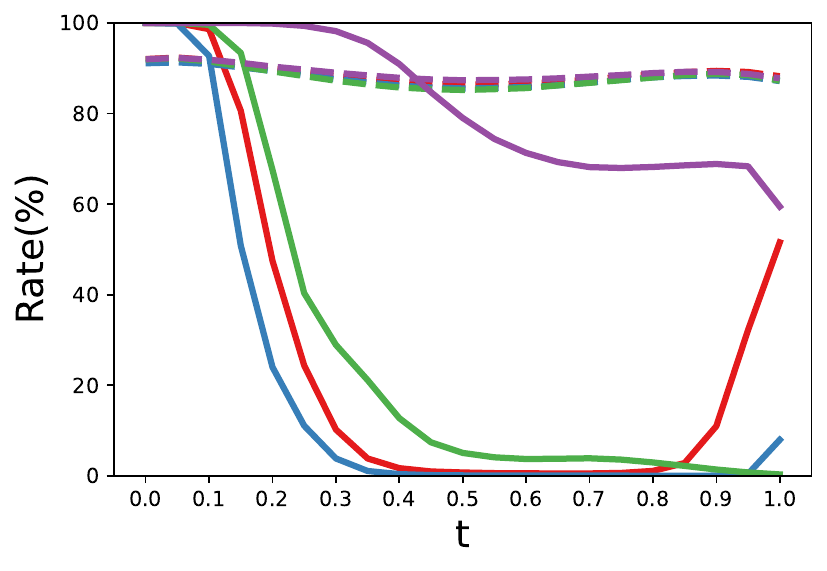}}
	\subfloat[\small{Unrelated}]{\includegraphics[height=3.35cm]{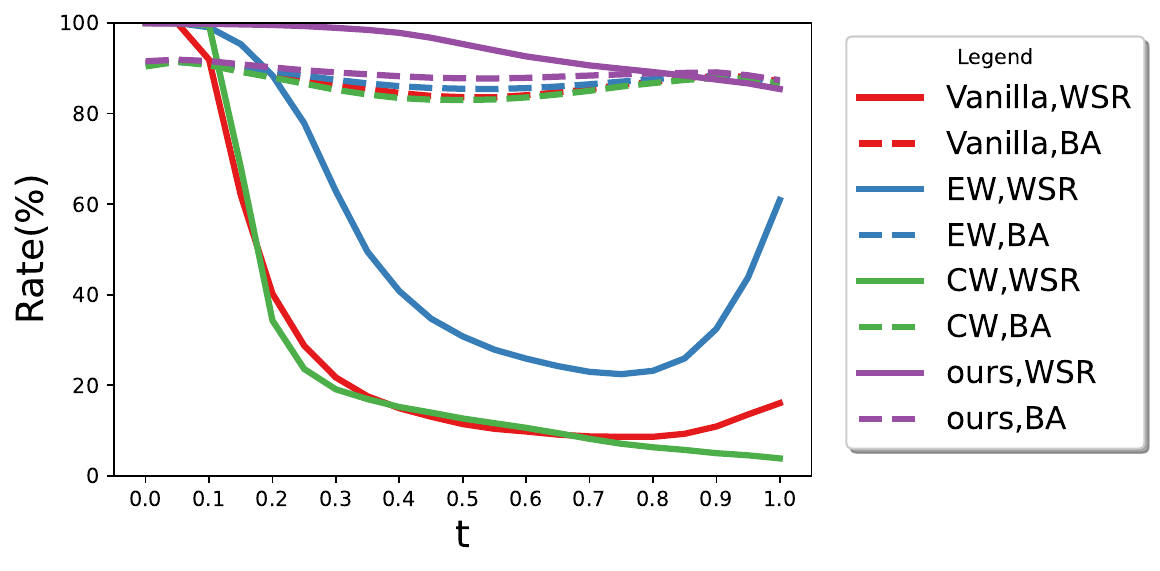}}
    \vspace{0.5em}
	\caption{MCR results with varying thresholds on CIFAR-10.}
	\label{fig:mcr-varying-threshold}
\end{figure*}

\begin{figure*}[!htbp]
    \centering
	\subfloat[\small{Content}]{\includegraphics[height=3.35cm]{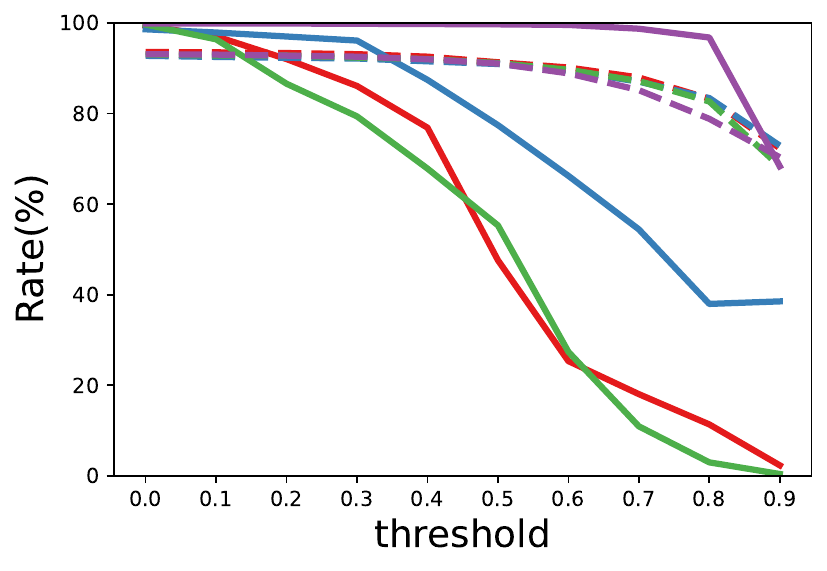}}
	\subfloat[\small{Noise}]{\includegraphics[height=3.35cm]{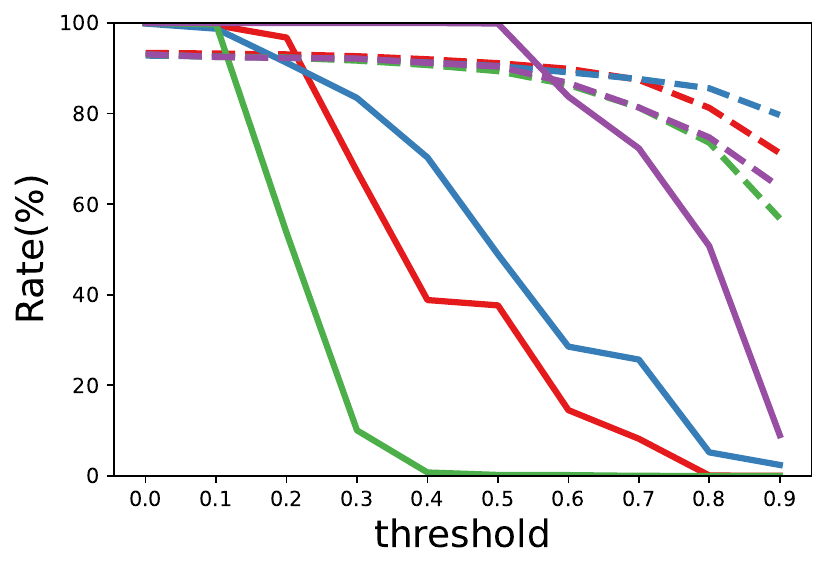}}
	\subfloat[\small{Unrelated}]{\includegraphics[height=3.35cm]{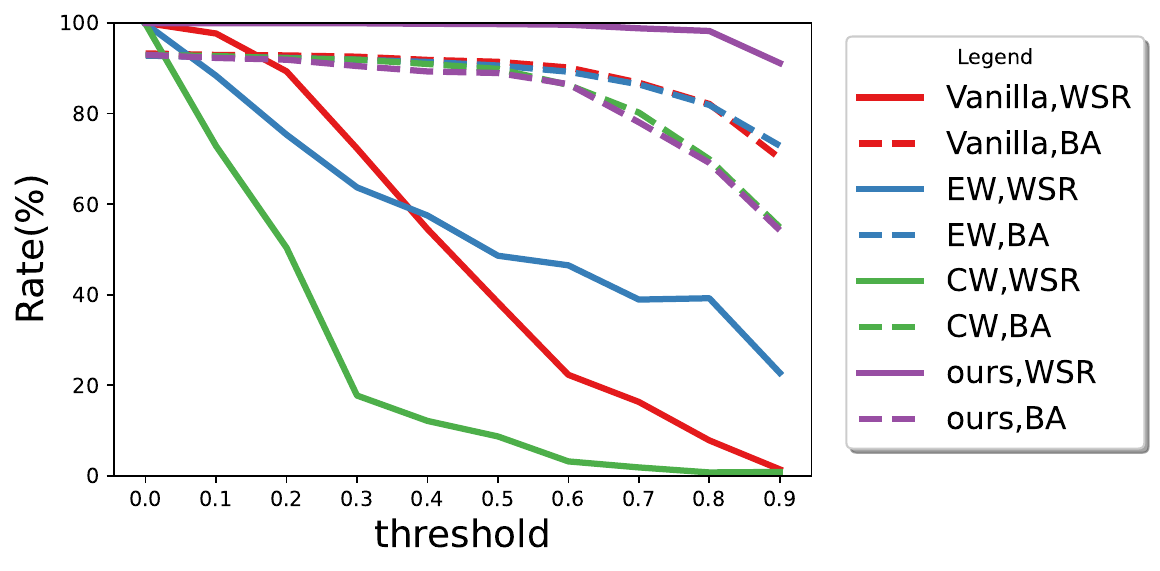}}
    \vspace{0.5em}
	\caption{ANP results with varying thresholds on CIFAR-10.}
	\label{fig:anp-varying-threshold}
\end{figure*}

%------------------------------------------------------------------------

\subsection{Detailed Results on CIFAR-100} \label{app:cifar-100}
To show that our method can also apply to other datasets, we conduct additional experiments on CIFAR-100.

\vspace{0.3em}\noindent \textbf{Modification to Attack Settings.} As trigger reconstruction need to scan 100 classes on CIFAR-100, we reduce the NC reconstruction epoch from 15 to 5 to speed it up. The ANP pruning threshold is set to 0.5 in CIFAR-100 experiments to maintain benign accuracy.

\vspace{0.3em}\noindent \textbf{Results.} As shown in Table~\ref{tab:cifar-100-full}, similar to previous results on CIFAR-10, our methods generally achieves better watermark robustness compared with other methods. The only exception is on noise watermark where all watermark embedding schemes failed to protect the watermark against FP and NNL attacks. Moreover, we can observe from Figure~\ref{fig:cifar100-mcr-varying-threshold} and Figure~\ref{fig:cifar100-anp-varying-threshold} that our models still outperform other methods regardless of the threshold value for ANP and MCR, in terms of robustness against them. 

% Table generated by Excel2LaTeX from sheet '100epoch_pr1e-2_cifar100'
\begin{table*}[!t]
  \centering
  \small
  \caption{Results on CIFAR-100. `NA' denotes `No Attack'.}
  % \vspace{-0.7em}
  \scalebox{0.98}{
    \begin{tabular}{ccccccccccc}
    \toprule
    Metric & Type & Method & NA    & FT    & FP    & ANP   & NAD   & MCR   & NNL & AvgDrop \\
    \midrule
    \multirow{12}[6]{*}{WSR} & \multirow{4}[2]{*}{Content} & Vanilla    & 98.27 & 19.63 & 1.96  & 70.25 & 0.62  & 15.14 & 0.24  & $\downarrow$ 80.30 \\
          &       & EW    & 97.93 & 10.57 & 2.84  & 55.23 & 1.92  & 1.44  & 1.14  & $\downarrow$ 85.74 \\
          &       & CW    & 98.77 & 11.80 & 0.23  & 12.12 & 0.44  & 11.65 & 0.09  & $\downarrow$ 92.72 \\
          &       & Ours  & \textbf{99.48} & \textbf{97.17} & \textbf{93.35} & \textbf{99.16} & \textbf{90.59} & \textbf{95.78} & \textbf{30.30} & $\downarrow$ \textbf{15.09} \\
\cmidrule(lr){2-11}          & \multirow{4}[2]{*}{Noise} & Vanilla    & 99.94 & 60.54 & \textbf{10.03} & 96.55 & 20.57 & 52.77 & 0.12 & $\downarrow$ 59.85 \\
          &       & EW    & 99.87 & 10.73 & 9.79  & 95.62 & 6.69  & 8.75  & \textbf{12.99} & $\downarrow$ 75.78 \\
          &       & CW    & 99.98 & 24.38 & 1.80  & 55.95 & 3.28  & 38.44 & 0.05 & $\downarrow$ 79.33 \\
          &       & Ours  & \textbf{100.00} & \textbf{84.82} & 8.60  & \textbf{99.99} & \textbf{73.67} & \textbf{93.82} & 0.98 & $\downarrow$ \textbf{39.69} \\
\cmidrule(lr){2-11}           & \multirow{4}[2]{*}{Unrelated} & Vanilla    & \textbf{100.00} & 6.83  & 1.50  & 92.25 & 6.25  & 12.58 & 11.42 & $\downarrow$ 78.19\\
          &       & EW    & \textbf{100.00} & 27.67 & 3.42  & 93.33 & 18.25 & 17.75 & 40.25 & $\downarrow$ 66.56 \\
          &       & CW    & 99.83 & 0.25  & 1.08  & 41.08 & 4.08  & 7.67  & 0.58 & $\downarrow$ 90.71\\
          &       & Ours  & \textbf{100.00} & \textbf{97.42} & \textbf{44.67} & \textbf{100.00} & \textbf{94.08} & \textbf{97.25} & \textbf{45.17} & $\downarrow$ \textbf{20.24} \\
    \midrule
    \midrule
    \multirow{12}[6]{*}{ACC} & \multirow{4}[2]{*}{Content} & Vanilla    & 73.78 & 69.43 & 68.34 & 67.83 & 65.86 & 63.72 & 66.40 & 6.85 \\
          &       & EW    & 73.45 & 67.91 & 66.76 & 66.33 & 63.69 & 61.22 & 66.93 & 7.97 \\
          &       & CW    & 73.95 & 68.98 & 68.42 & 61.97 & 65.06 & 63.25 & 67.92 & 8.01 \\
          &       & Ours  & 73.35 & 68.86 & 67.99 & 68.07 & 65.86 & 63.95 & 67.89 & 6.25 \\
\cmidrule(lr){2-11}          & \multirow{4}[2]{*}{Noise} & Vanilla    & 74.13 & 69.61 & 68.78 & 70.72 & 66.30 & 63.73 & 67.30 & 6.39 \\
          &       & EW    & 73.43 & 67.39 & 66.92 & 68.85 & 64.18 & 61.10 & 66.96 & 7.53 \\
          &       & CW    & 73.49 & 68.00 & 67.84 & 59.21 & 64.26 & 61.68 & 66.79 &  8.86 \\
          &       & Ours  & 72.97 & 68.49 & 67.39 & 67.59 & 64.94 & 63.08 & 66.25 & 6.68 \\
\cmidrule(lr){2-11} 
& \multirow{4}[2]{*}{Unrelated} & Vanilla    & 73.80 & 68.55 & 67.46 & 69.90 & 65.14 & 61.87 & 65.77 & 7.35\\
          &       & EW    & 73.57 & 67.83 & 66.61 & 69.39 & 63.52 & 61.47 & 65.90 & 7.78 \\
          &       & CW    & 73.45 & 67.45 & 66.90 & 54.59 & 62.66 & 60.60 & 64.88 & 10.60 \\
          &       & Ours  & 72.27 & 67.68 & 66.88 & 65.22 & 64.07 & 61.99 & 62.64 & 7.53 \\
    \bottomrule
    \end{tabular}%
    }
  \label{tab:cifar-100-full}%
\end{table*}%

\begin{figure*}[!htbp]
    \centering
	\subfloat[\small{Content}]{\includegraphics[height=3.35cm]{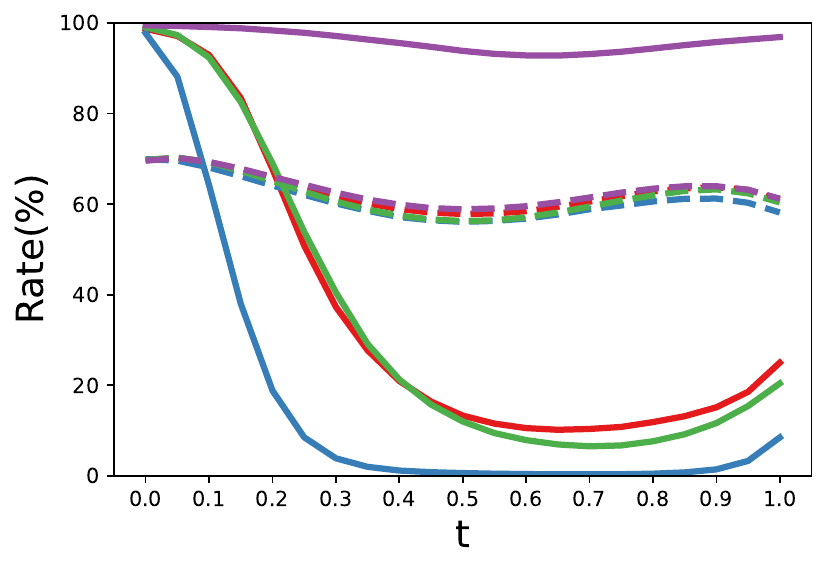}}
	\subfloat[\small{Noise}]{\includegraphics[height=3.35cm]{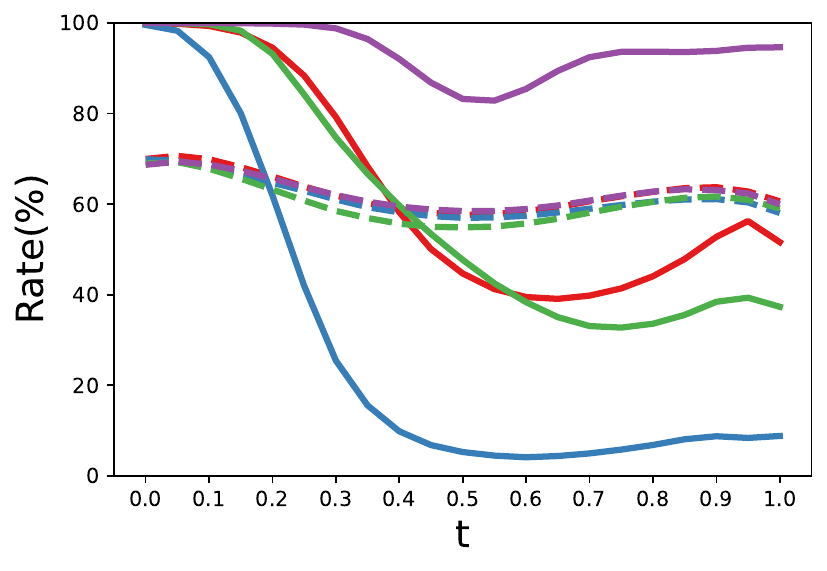}}
	\subfloat[\small{Unrelated}]{\includegraphics[height=3.35cm]{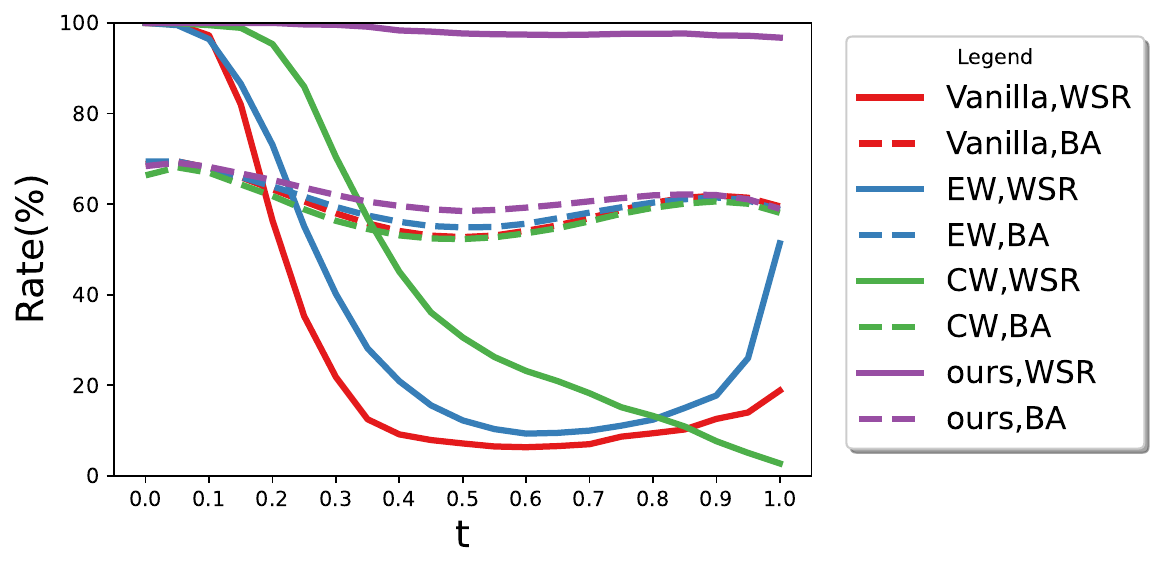}}
	\vspace{0.5em}
	\caption{MCR results with varying thresholds on CIFAR-100.}
	\label{fig:cifar100-mcr-varying-threshold}
\end{figure*}

\begin{figure*}[!htbp]
    \centering
	\subfloat[\small{Content}]{\includegraphics[height=3.35cm]{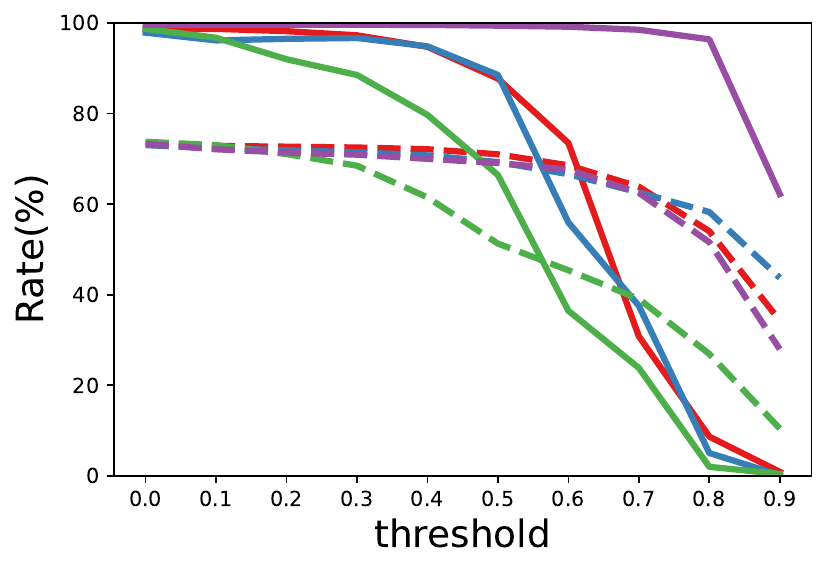}}
	\subfloat[\small{Noise}]{\includegraphics[height=3.35cm]{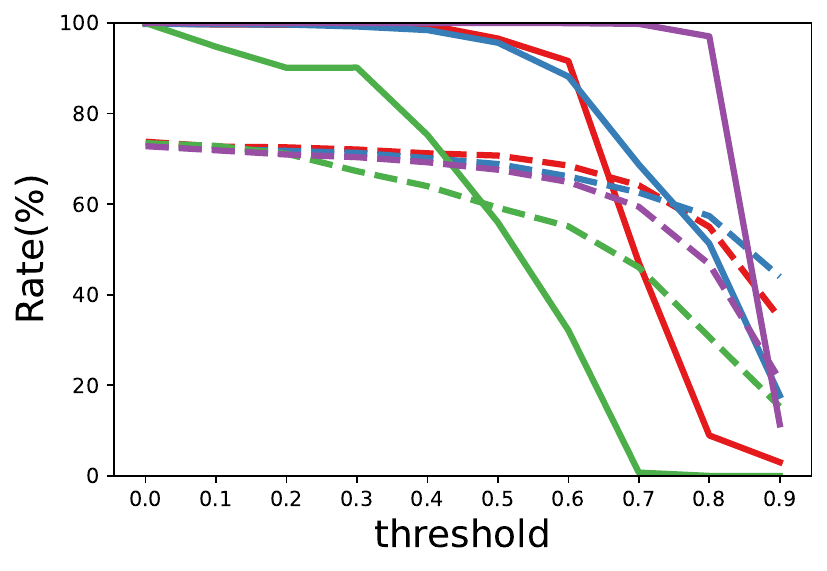}}
	\subfloat[\small{Unrelated}]{\includegraphics[height=3.35cm]{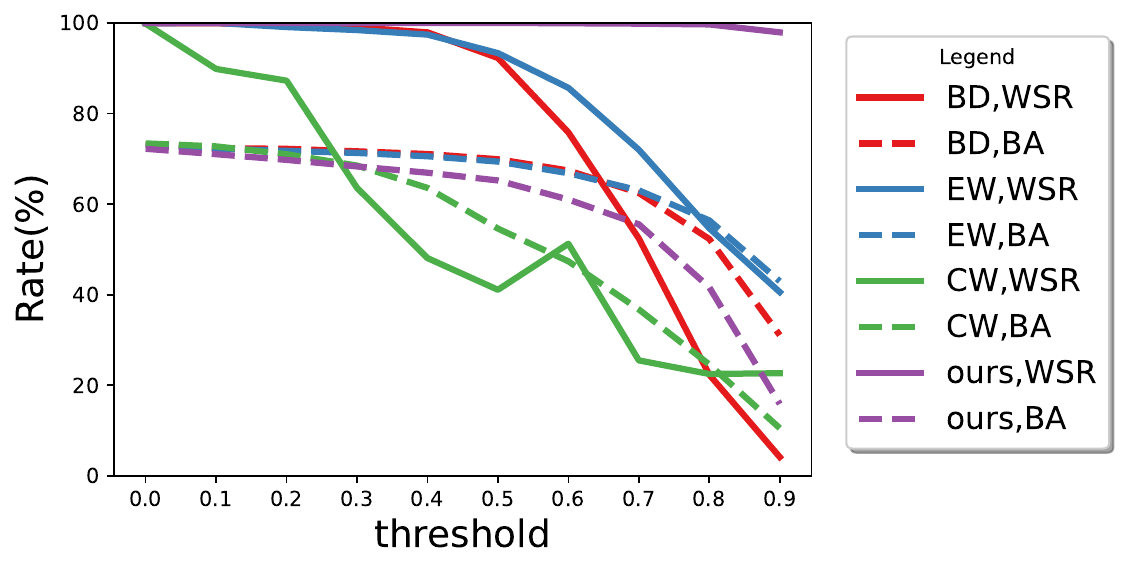}}
        \vspace{0.5em}
	\caption{ANP results with varying thresholds on CIFAR-100.}
	\label{fig:cifar100-anp-varying-threshold}
\end{figure*}

\subsection{Detailed Results on ImageNet Subset} \label{app:imn-100}
To verify that our model can apply to other datasets, we experiment on a subset of ImageNet, containing 100 classes with 50,000 images for training (500 images per class) and 5,000 images for testing (50 images per class). , and the results are shown as follows.

\vspace{0.3em}\noindent \textbf{Modification to Defense Settings.} We set the perturbation budget $\epsilon$ to $1\times10^{-3}$ for better benign accuracy.

% Table generated by Excel2LaTeX from sheet '100epoch_pr1e-2_cifar100'
\begin{table*}[!t]
  \centering
  \small
  \caption{Results on ImageNet subset. `NA' denotes `No Attack'.}
  % \vspace{-0.7em}
  \scalebox{0.98}{
    \begin{tabular}{ccccccccccc}
    \toprule
    Metric & Type & Method & NA    & FT    & FP    & ANP   & NAD   & MCR   & NNL & AvgDrop \\
\midrule
    \multirow{12}[6]{*}{WSR} & \multirow{4}[2]{*}{Content} & Vanilla & 98.26 & 22.18 & 9.31  & 43.91 & 4.40  & 12.48 & 28.05 & $\downarrow$ 78.20 \\
          &       & EW    & 95.85 & 8.95  & 3.82  & 17.07 & 3.02  & 8.82  & 19.96 & $\downarrow$ 85.58 \\
          &       & CW    & 99.05 & 6.35  & 0.16  & 0.26  & 0.68  & 2.92  & 17.91 & $\downarrow$ 94.34 \\
          &       & Ours  & \textbf{99.54} & \textbf{57.56} & \textbf{21.46} & \textbf{98.57} & \textbf{31.95} & \textbf{71.93} & \textbf{79.39} & $\downarrow$ \textbf{39.40} \\
\cmidrule{2-11}          & \multirow{4}[2]{*}{Noise} & Vanilla & 98.65 & 9.54  & 2.79  & 29.00 & 9.75  & 8.06  & \textbf{3.60} & $\downarrow$ $\downarrow$ 88.20 \\
          &       & EW    & 95.36 & 3.58  & 4.08  & 1.19  & 1.62  & 4.19  & 1.56  & $\downarrow$ 92.66 \\
          &       & CW    & 98.32 & 15.35 & 2.57  & 11.65 & 5.65  & 3.41  & 2.56  & $\downarrow$ 91.45 \\
          &       & Ours  & \textbf{99.38} & \textbf{33.80} & \textbf{11.69} & \textbf{95.52} & \textbf{32.54} & \textbf{28.40} & 1.43  & $\downarrow$ \textbf{65.48} \\
\cmidrule{2-11}          & \multirow{4}[2]{*}{Unrelated} & Vanilla & 99.97 & 47.40 & 36.53 & 99.66 & 24.16 & 54.43 & 30.87 & $\downarrow$ 51.13 \\
          &       & EW    & 99.97 & 33.64 & 31.12 & 94.40 & 59.91 & 12.94 & 56.70 & $\downarrow$ 51.85 \\
          &       & CW    & 99.99 & 38.94 & 0.86  & 1.97  & 43.68 & 65.74 & 26.66 & $\downarrow$ 70.34 \\
          &       & Ours  & \textbf{100.00} & \textbf{93.98} & \textbf{81.97} & \textbf{99.99} & \textbf{88.99} & \textbf{93.97} & \textbf{96.57} & $\downarrow$ \textbf{7.42} \\
    \midrule
    \midrule
    \multirow{12}[6]{*}{BA} & \multirow{4}[2]{*}{Content} & Vanilla & 74.81 & 69.88 & 70.37 & 65.76 & 66.17 & 67.90 & 69.75 & 6.50 \\
          &       & EW    & \textbf{75.15} & 68.66 & 69.18 & 61.91 & 64.15 & 65.65 & 69.42 & 8.65 \\
          &       & CW    & 74.52 & 69.67 & 70.02 & 51.55 & 65.70 & 66.30 & 69.16 & 9.12 \\
          &       & Ours  & 72.29 & 68.37 & 68.35 & 56.21 & 64.55 & 66.21 & 66.53 & 7.26 \\
\cmidrule{2-11}          & \multirow{4}[2]{*}{Noise} & Vanilla & 74.47 & 70.05 & 70.63 & 65.77 & 67.23 & 67.50 & 71.11 & 5.76 \\
          &       & EW    & \textbf{75.09} & 68.06 & 69.51 & 60.96 & 64.16 & 65.51 & 69.53 & 8.80 \\
          &       & CW    & 74.11 & 69.37 & 70.09 & 54.15 & 65.34 & 66.63 & 70.87 & 8.03 \\
          &       & Ours  & 71.48 & 67.25 & 67.45 & 30.74 & 62.60 & 63.52 & 58.71 & 13.10 \\
\cmidrule{2-11}          & \multirow{4}[2]{*}{Unrelated} & Vanilla & 74.69 & 69.92 & 70.57 & 65.77 & 66.79 & 67.45 & 70.13 & 6.25 \\
          &       & EW    & \textbf{75.25} & 68.38 & 69.32 & 60.63 & 64.67 & 65.95 & 70.01 & 8.76 \\
          &       & CW    & 74.97 & 70.05 & 70.81 & 54.05 & 66.31 & 66.89 & 70.13 & 8.60 \\
          &       & Ours  & 73.55 & 68.97 & 69.63 & 57.41 & 64.96 & 66.93 & 68.69 & 7.45 \\
    \bottomrule

    \end{tabular}%
    }
  \label{tab:imn-100-full}%
\end{table*}%

\begin{figure*}[!htbp]
    \centering
	\subfloat[\small{Content}]{\includegraphics[height=3.35cm]{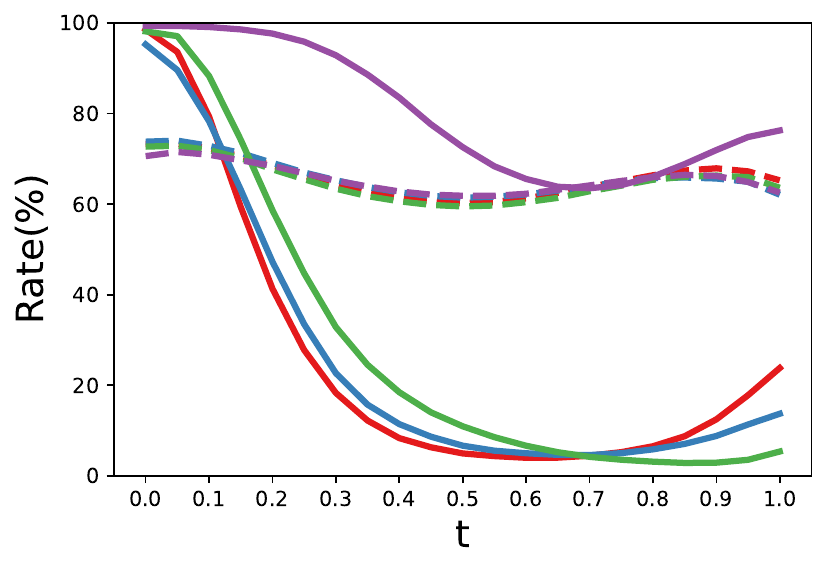}}
	\subfloat[\small{Noise}]{\includegraphics[height=3.35cm]{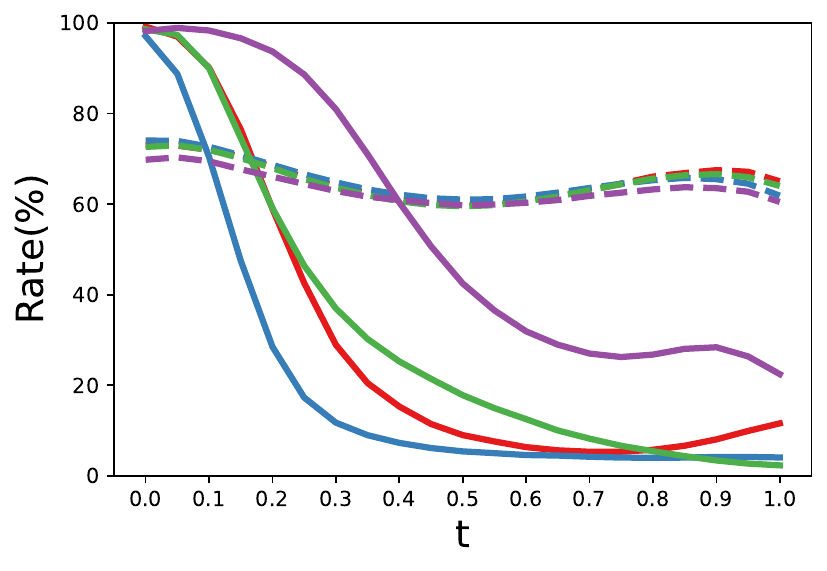}}
	\subfloat[\small{Unrelated}]{\includegraphics[height=3.35cm]{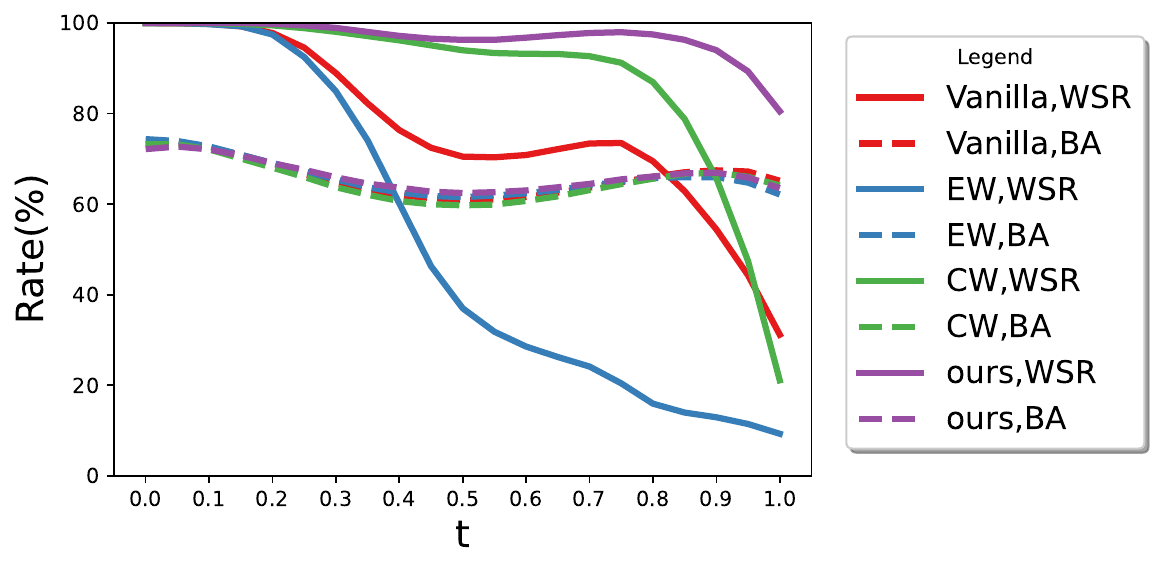}}
        \vspace{0.5em}
	\caption{MCR results with varying thresholds on ImageNet subset.}
	\label{fig:imn100-mcr-varying-threshold}
\end{figure*}

\begin{figure*}[!htbp]
    \centering
	\subfloat[\small{Content}]{\includegraphics[height=3.35cm]{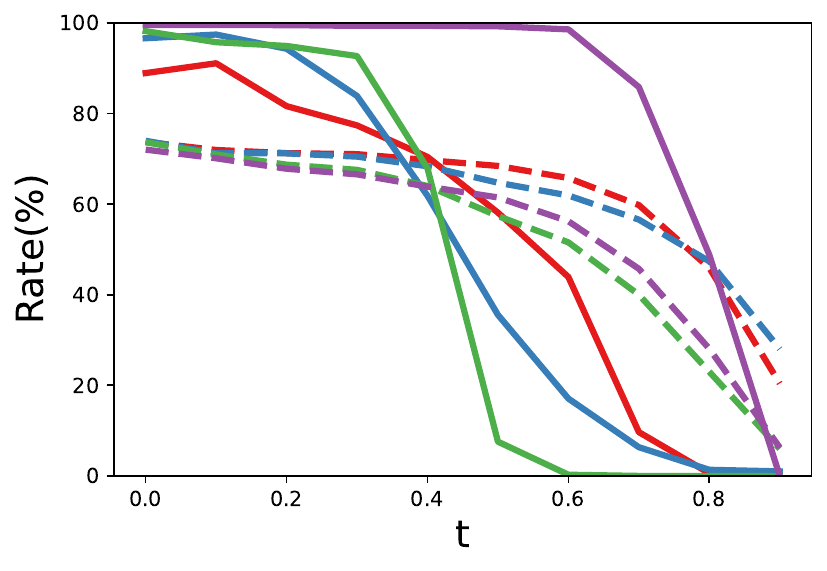}}
	\subfloat[\small{Noise}]{\includegraphics[height=3.35cm]{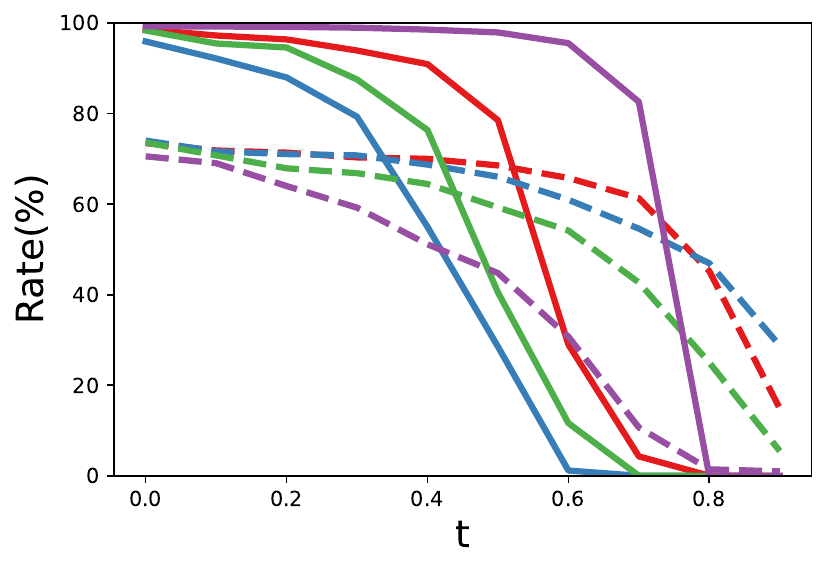}}
	\subfloat[\small{Unrelated}]{\includegraphics[height=3.35cm]{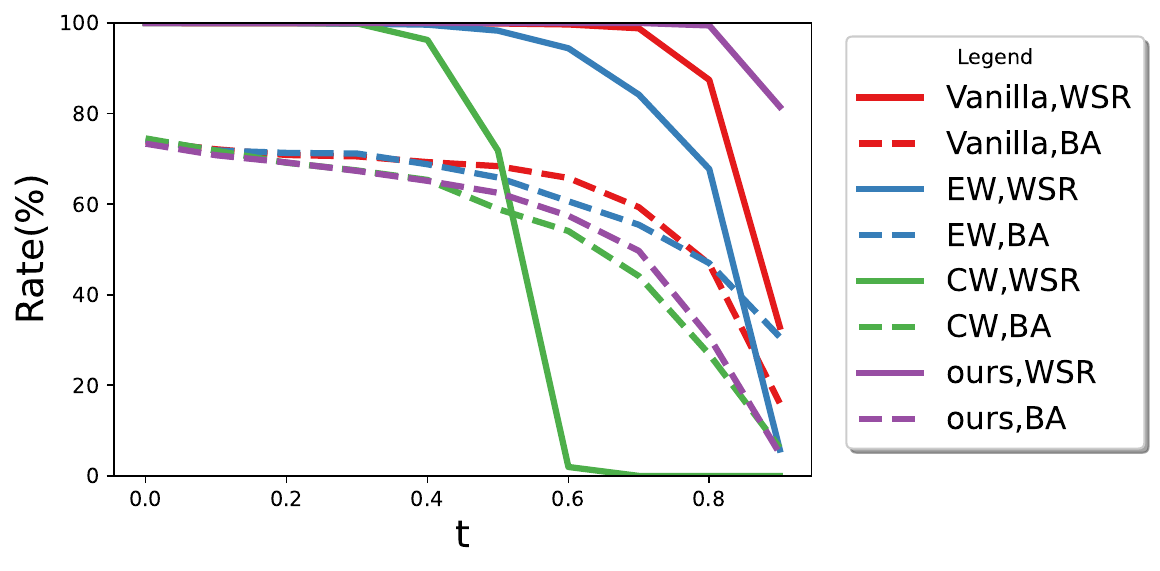}}
	\caption{ANP results with varying thresholds on ImageNet subset.}
	\label{fig:imn100-anp-varying-threshold}
 \vspace{0.5em}
\end{figure*}

\begin{table*}[!ht]
  \centering
  \small
  \caption{Results of Content embedded with varying perturbation magnitude $\epsilon$ using our method. AVG denotes the average WSR/BA after watermark removal attacks.}
  % \vspace{-0.7em}
    \begin{tabular}{cccccccccc}
    \toprule
    Metric & $\epsilon$   & NA    & FT    & FP    & ANP   & NAD   & MCR   & NNL   & AVG \\
    \midrule
    \multirow{5}[2]{*}{WSR} & $5\times10^{-3}$ & 99.86 & 88.93 & 93.31 & 96.61 & 61.94 & 39.04 & 44.43 & 70.71 \\
          & $1\times10^{-2}$ & 99.87 & 95.57 & 97.00 & 99.03 & 63.40 & 59.85 & 65.20 & 80.01 \\
          & $2\times10^{-2}$ & 99.87 & \textbf{96.63} & \textbf{98.44} & 99.56 & \textbf{90.76} & 84.65 & \textbf{68.58} & \textbf{89.77} \\
          & $4\times10^{-2}$ & \textbf{99.90} & 95.88 & 97.25 & \textbf{99.64} & 81.50 & \textbf{89.94} & 66.57 & 88.46 \\
          % & $8\times10^{-2}$ & 99.76 & 84.24 & 80.51 & 97.40 & 53.90 & 67.18 & 18.17 & 67.58 \\
    \midrule
    \multirow{5}[2]{*}{BA} & $5\times10^{-3}$ & 93.50 & 91.90 & 91.96 & 89.53 & 90.06 & 89.30 & 91.43 & 90.70 \\
          & $1\times10^{-2}$ & 93.62 & 91.76 & 92.18 & 89.50 & 90.40 & 89.15 & 91.63 & 90.77 \\
          & $2\times10^{-2}$ & 93.42 & 91.72 & 91.81 & 88.86 & 89.79 & 89.08 & 91.06 & 90.39 \\
          & $4\times10^{-2}$ & 93.34 & 91.51 & 91.85 & 87.15 & 89.63 & 89.16 & 90.67 & 89.99 \\
          % & $8\times10^{-2}$ & 93.14 & 91.35 & 91.48 & 86.91 & 89.59 & 89.04 & 90.24 & 90.14 \\
    \bottomrule
    \end{tabular}%
  \label{tab:varying-eps}%
\end{table*}%

\vspace{0.3em}\noindent \textbf{Modification to Attack Settings.} As trigger reconstruction need to scan 100 classes on the ImageNet subset, we reduce the NC reconstruction epoch from 15 to 5 to speed it up.

\vspace{0.3em}\noindent \textbf{Results.} As shown in Table~\ref{tab:imn-100-full}, similar to previous results on CIFAR-10, our methods generally reaches better watermark robustness compared with other methods. The only exception is on noise watermark, where all watermark embedding schemes failed to protect the watermark against NNL attacks. Moreover, we can observe from Figure~\ref{fig:imn100-mcr-varying-threshold} and Figure~\ref{fig:imn100-anp-varying-threshold} that our models still outperform other methods regardless of the threshold value for ANP and MCR, in terms of robustness against them.

%------------------------------------------------------------------------
\section{Detailed Results of Ablation Studies}
\subsection{Results with Varying Perturbation Magnitude} \label{app:varying-eps}
We visualize some results of the Content watermark embedded with different perturbation magnitude $\epsilon$ in Sec~\ref{subsec:ablation-study}. Here, we provided more detailed results in a numeric form in Table~\ref{tab:varying-eps}. Generally speaking, our method consistently improves the robustness of the watermark, with the watermark success rate higher than other methods throughout all tested $\epsilon$. Moreover, the amount of improvement against all evaluated attacks shows similar trends, and this consistent robustness improvement benefits the selection of perturbation magnitude $\epsilon$.

\subsection{Results with Other Target Classes} \label{app:varying-target}
To demonstrate that our method can apply to different target classes, we experimented with Content and set the target class $y_t \in \{1,2,3,4\}$. Similar to the default scenario where $y_t=0$, these 4 tests maintain the average watermark success rate of $94.87\%$, $79.81\%$, $84.36\%$ and $87.76\%$ respectively under all 6 removal attacks, while the standard baseline only achieves $32.91\%, 20.79\%, 32.28\%$, and $10.13\%$ against the above six attacks, indicating that our method achieves stable robustness improvement regardless of the chosen target class (as shown in Table~\ref{tab:varying-yt-bd}-\ref{tab:varying-yt-ours}).

% Table generated by Excel2LaTeX from sheet '100epoch_pr1e-2_cifar10_bd_vary'
\begin{table*}[!]
  \centering
  \small
  \caption{Results of vanilla model watermark over content-type attack with different target labels.}
  % \vspace{-0.7em}
  \vspace{0.5em}
    \begin{tabular}{cccccccccc}
    \toprule
    Metric &  $y_t$ & NA    & FT    & FP    & ANP   & NAD   & MCR   & NNL   & AVG \\
    \midrule
    \multirow{5}[2]{*}{WSR} & 0     & 99.56 & 56.78 & 74.58 & 25.34 & 48.14 & 16.56 & 21.02 & 40.40 \\
          & 1     & 99.51 & 46.54 & 73.60 & 45.93 & 12.83 & 9.41  & 9.15  & 32.91 \\
          & 2     & 99.54 & 47.97 & 55.16 & 9.24  & 3.23  & 6.61  & 2.52  & 20.79 \\
          & 3     & 99.48 & 60.79 & 77.99 & 8.56  & 15.89 & 11.87 & 18.56 & 32.28 \\
          & 4     & 99.53 & 17.13 & 10.39 & 9.07  & 11.39 & 8.50  & 4.33  & 10.13 \\
    \midrule
    \multirow{5}[2]{*}{BA} & 0     & 93.86 & 91.80 & 92.19 & 90.15 & 90.39 & 89.27 & 91.92 & 90.95 \\
          & 1     & 93.85 & 92.27 & 92.31 & 90.03 & 90.38 & 89.39 & 91.87 & 91.04 \\
          & 2     & 93.61 & 91.74 & 92.01 & 89.60 & 90.16 & 88.87 & 91.67 & 90.67 \\
          & 3     & 93.90 & 92.01 & 92.11 & 90.77 & 90.07 & 89.26 & 92.04 & 91.04 \\
          & 4     & 93.85 & 91.93 & 92.20 & 90.64 & 90.34 & 89.23 & 91.52 & 90.98 \\
    \bottomrule
    \end{tabular}%
  \label{tab:varying-yt-bd}%
\end{table*}%

\begin{table*}[!htbp]
  \centering
  \small
  \caption{Results of our model watermark over content-type attack with different target labels.}
  \vspace{0.5em}
    \begin{tabular}{cccccccccc}
    \toprule
    Metric & $y_t$ & NA    & FT    & FP    & ANP   & NAD   & MCR   & NNL   & AVG \\
    \midrule
    \multirow{5}[2]{*}{WSR} & 0     & 99.87 & 96.63 & 98.44 & 99.56 & 90.76 & 84.65 & 68.58 & 89.77 \\
          & 1     & 99.76 & 97.69 & 98.58 & 99.49 & 90.20 & 90.12 & 93.16 & 94.87 \\
          & 2     & 99.76 & 95.60 & 97.55 & 98.95 & 53.68 & 73.04 & 60.03 & 79.81 \\
          & 3     & 99.76 & 97.30 & 97.22 & 98.83 & 65.81 & 82.96 & 64.01 & 84.36 \\
          & 4     & 99.73 & 97.02 & 97.31 & 99.13 & 78.91 & 76.84 & 77.35 & 87.76 \\
    \midrule
    \multirow{5}[2]{*}{BA} & 0     & 93.42 & 91.72 & 91.81 & 88.86 & 89.79 & 89.08 & 91.06 & 90.39 \\
          & 1     & 93.63 & 91.58 & 92.09 & 89.61 & 90.19 & 89.03 & 91.50 & 90.67 \\
          & 2     & 93.31 & 91.59 & 91.71 & 88.67 & 89.72 & 88.80 & 91.23 & 90.29 \\
          & 3     & 93.73 & 91.69 & 91.67 & 89.29 & 89.92 & 89.05 & 91.16 & 90.46 \\
          & 4     & 93.38 & 91.50 & 91.88 & 85.58 & 89.46 & 89.05 & 91.13 & 89.77 \\
    \bottomrule
    \end{tabular}%
  \label{tab:varying-yt-ours}%
\end{table*}%

%------------------------------------------------------------------------
%------------------------------------------------------------------------

\section{Additional Ablation Experiments}

\subsection{Results with other model architectures}
In Section~\ref{subsec:ablation-study}, we demonstrate that our method improves watermark robustness against the FT attack across various model architectures ($i.e.$, MobileNetV2, VGG16, and ResNet50). To further verify that our method is better than baseline defenses across different model architectures under different attacks, in this section, we conduct additional experiments under more attacks ($i.e.$, ANP, NAD, MCR) other than FT-based attacks. As shown in Figure~\ref{fig:more-attack-results}, our method consistently improves the watermark robustness across different model architectures under all attacks.

\begin{table*}[!htb]
  \centering
  \small
  \caption{The results with MobileNetV2 on CIFAR-10.}
  \vspace{0.5em}
  \scalebox{1.}{
    \begin{tabular}{cc|cccccccc}
    \toprule
    Metric & Method & NA    & FT    & FP    & ANP   & NAD   & MCR   & NNL   & AvgDrop \\
    \midrule
    \multirow{4}[0]{*}{WSR} & Vanilla & 99.07 & 24.44 & 44.14 & 77.15 & 32.86 & 23.09 & 19.44 & $\downarrow$ 62.22 \\
          & EW    & 98.59 & 15.96 & 9.47  & 63.56 & 23.58 & 11.66 & 13.71 & $\downarrow$ 75.60 \\
          & CW    & 99.16 & 34.32 & 23.75 & 29.01 & 26.66 & 15.28 & 20.62 & $\downarrow$ 74.22 \\
          & Ours  & \textbf{99.77} & \textbf{67.84} & \textbf{66.78} & \textbf{99.94} & \textbf{82.73} & \textbf{48.73} & \textbf{53.06} & $\downarrow$ \textbf{29.93} \\
    \midrule
    \multirow{4}[0]{*}{BA} & Vanilla & \textbf{92.27} & 89.28 & 90.15 & 68.50 & 87.56 & 85.52 & 89.08 & 7.26 \\
          & EW    & 90.04 & 86.41 & 87.38 & 84.83 & 82.59 & 79.49 & 87.54 & 5.33 \\
          & CW    & 92.07 & 89.06 & 89.38 & 77.70 & 86.88 & 85.20 & 88.97 & 5.87 \\
          & Ours  & 90.99 & 88.33 & 88.06 & 57.07 & 85.51 & 83.10 & 87.99 & 9.32 \\
    \bottomrule
    \end{tabular}%
    }
  \label{tab:mobilenetv2}%
\end{table*}%

\begin{figure*}[!htb]
% 	\centering
    \centering
	\subfloat[{ANP}]{\includegraphics[height=3.75cm]{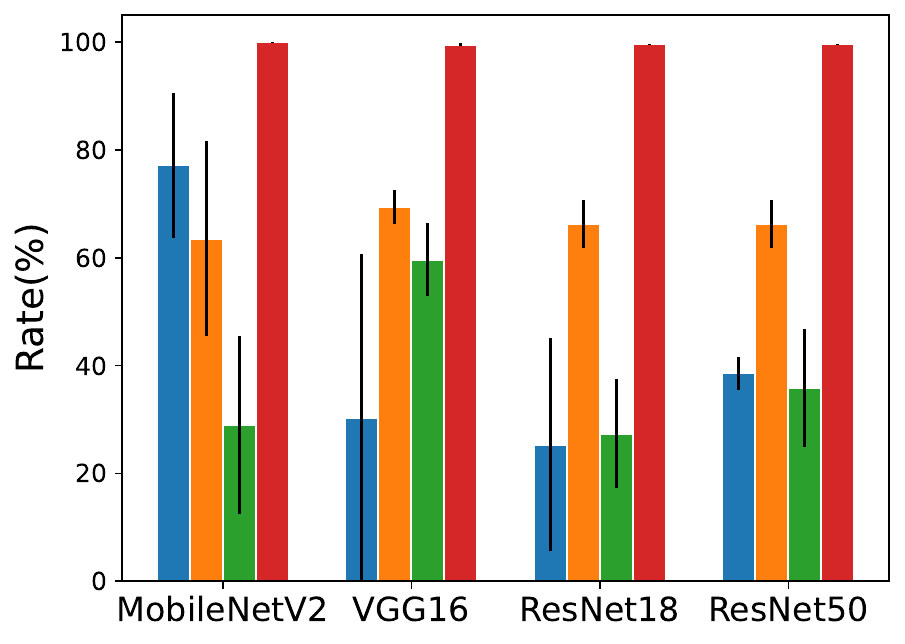}}
    % \hspace*{-3em}
	\subfloat[{NAD}]{\includegraphics[height=3.75cm]{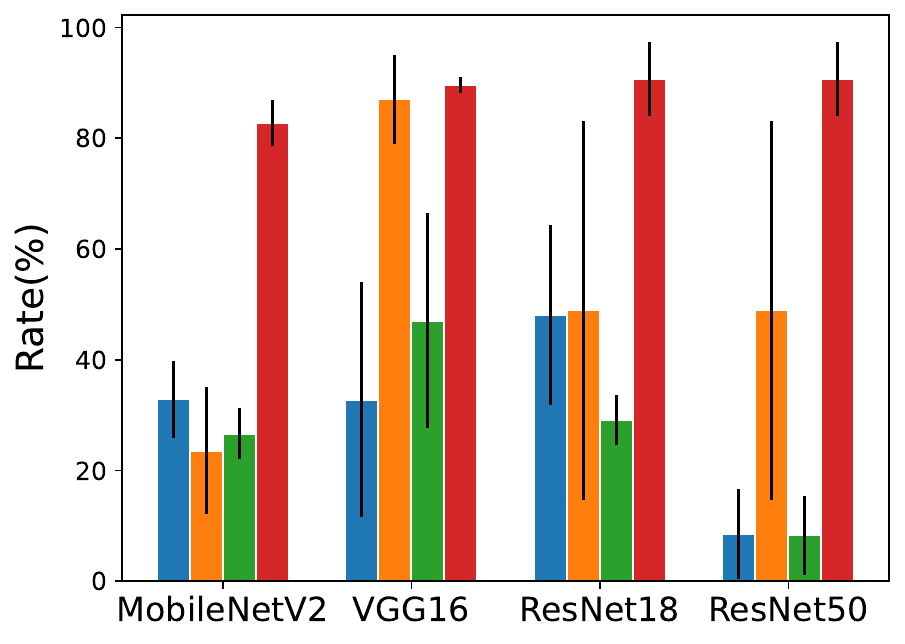}}
	\subfloat[{MCR}]{\includegraphics[height=3.75cm]{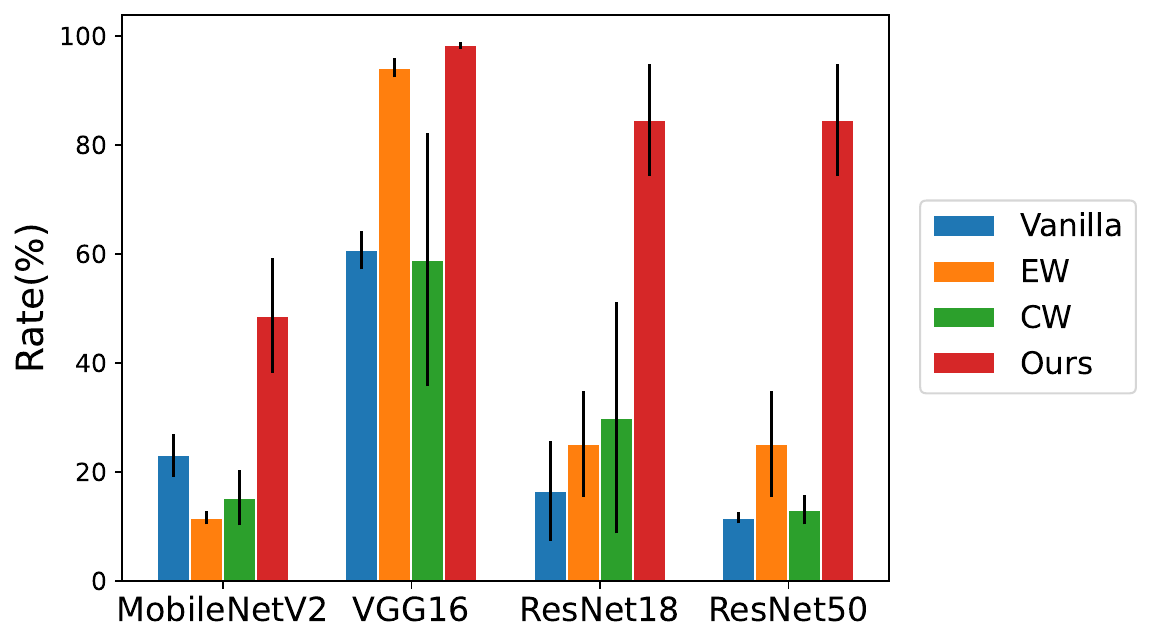}}
    % \vspace{-0.6em}
	\caption{The WSR of models under ANP, NAD, and MCR.}
	\label{fig:more-attack-results}
\end{figure*}

In addition, to further verify that our method is still effective under simpler model architecture, we conduct additional experiments on CIFAR-10 with MobileNetV2. MobileNetV2 consists of 2.2M trainable parameters, which is significantly less than the 11.2M parameters contained in ResNet18 used in our main experiments. As shown in Table~\ref{tab:mobilenetv2}, in this case, our method is still better than all baseline methods with the average WSR drop of 29.93\%, whereas all baseline defenses suffer from at least 62.22\% average WSR decreases. These results verify the effectiveness of our method again.
%------------------------------------------------------------------------
%------------------------------------------------------------------------

\section{Additional Robustness Evaluations}

\subsection{Comparison with Other Watermark Methods}
We compare our method with three other SOTA methods: NTL~\cite{wang2021non}, ROSE~\cite{kallas2022rose}, and CAE~\cite{lukas2020deep}. NTL uses the error rate on patched data to indicate WSR, \textit{i.e.}, the higher error rate is, the larger WSR is. While NTL lists ACC in the original paper, we list the error rate (= 1 - ACC) for easier comparison. The results are shown in Table~\ref{tab:ntl-rose}
Note that we apply a larger $lr$ for FT, which makes the defense more challenging. For fairness, we compare different methods with various $lr$ in the table including the results from original papers. Ours outperforms the others in almost all cases.
\begin{table}
    \vspace*{-0.2ex}
    % \hspace*{-1.5em}
    % \centering
    % \hspace{em}
    \small
    \caption{Results under FT attack with different learning rates. ``$*$" denotes results from the original paper. ``-" denotes results that are not reported in the original paper.}
    
    \scalebox{1.}{
    \begin{tabular}{c|c|cccc}
    \toprule
    Method & Before & 1e-5 & 1e-3 & 1e-2 & 2e-2 \\
    \midrule
    ROSE$^*$   & 92.50 & 92.50  & -- & - & -- \\
    ROSE   & 97.50 & 97.50  & 77.50 & 42.50 & 10.00 \\
    NTL$^*$ & 85.20 & --  & 86.50 & -- & -- \\
    NTL   & 87.51 & 89.69  & 89.45 & 46.42 & 36.71\\
    CAE$^*$ & 100.00 & --  & 100.00 & -- & -- \\
    CAE   & 100.00 & 100.00 & 100.00 & 94.67 & 81.00\\
    Ours  & 99.87 & 99.71 & 99.85 & 99.71  & 99.45  \\
    \bottomrule    
    \end{tabular}%
    }
  \label{tab:ntl-rose}%  
  \vspace*{-1.2em}
\end{table}
\vspace{-1.5em}
\subsection{Comparison with Adversarial Training}
% \vspace{-1.5em}
Some may wonder if input perturbation helps embody a more robust watermark. In general, adversarial training can increase the stability of model predictions to image perturbations. However, a robust watermark requires that the prediction is stable regarding the changes in model parameters (caused by watermark-removal attacks). Thus, AT does not necessarily improve the robustness of model watermarks. As shown in Table~\ref{tab:AT}, AT may even reduce watermark robustness. We will explore its mechanisms in the future.  

\begin{table}[hp]
\vspace{-0.3cm}
  \centering
  % \caption{\footnotesize{Results on CIFAR-10 with ``TEST" trigger.}}
  \caption{Comparison with AT methods.}
  \scalebox{0.75}{
    \begin{tabular}{c|c|cccccc}
    \toprule

    Method   & Before & FT    & FP    & ANP   & NAD   & MCR   & NNL  \\
    \midrule
    \textit{Vanilla} & 99.56 & 56.78 & 74.58 & 25.34 & 48.14 & 16.56 & 21.02  \\
    PGD-AT~\cite{madry2018towards} & 98.66 & 20.59 & 30.80 & 46.47 & 14.26 & 14.69 & 45.56  \\
    TRADES~\cite{zhang2019theoretically} & 98.97 & 46.24 & 37.42 & 23.77 & 5.45  & 15.47 & 54.67  \\    
    Ours & \textbf{99.87} & \textbf{96.63} & \textbf{98.44} & \textbf{99.56} & \textbf{90.76} & \textbf{84.65} & \textbf{68.58} \\
    \bottomrule
    \end{tabular}%
    }
    
  \label{tab:AT}%
\end{table}%

\section{Visualizing the Feature Space}\label{app:t-SNE}
To provide further understandings about the effectiveness of our method, we visualize the how the hidden representation evolves along the adversarial direction and during the process of fine-tuning via t-SNE~\cite{van2008visualizing}.
%------------------------------------------------------------------------
\subsection{Features Along with the Adversarial Direction}\label{app:tsne-adv}
To show how the hidden representation evolves along the adversarial direction, we add a small adversarial perturbation to the watermarked model with the perturbation magnitude growing by $2\times10^{-3}$ every step. As can see in Figure \ref{fig:tsne-adv-bd}-\ref{fig:tsne-adv-cw}, the representation of watermark samples quickly mixes with the clean representation under small perturbation. In contrast, our method manages to maintain the watermark samples in a distinct cluster and the cluster remains distant from the untargeted clusters, as shown in Figure \ref{fig:tsne-adv-ours}.
\begin{figure*}[t]
	\centering
	\subfloat[]{\includegraphics[width=0.18\linewidth]{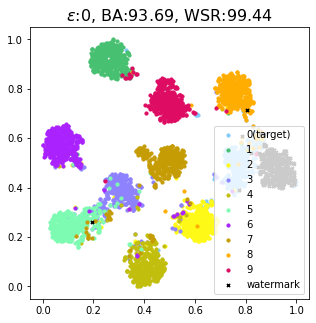}}
	\subfloat[]{\includegraphics[width=0.18\linewidth]{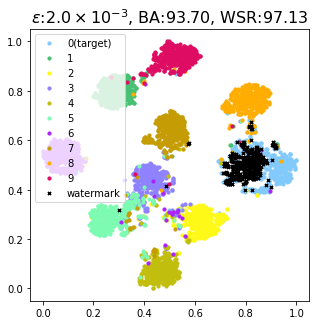}}
	\subfloat[]{\includegraphics[width=0.18\linewidth]{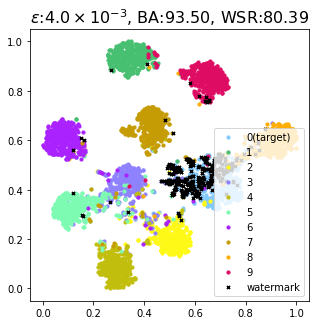}}
	\subfloat[]{\includegraphics[width=0.18\linewidth]{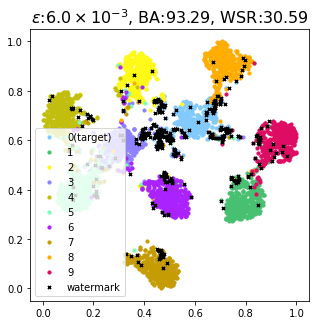}}
	\subfloat[]{\includegraphics[width=0.18\linewidth]{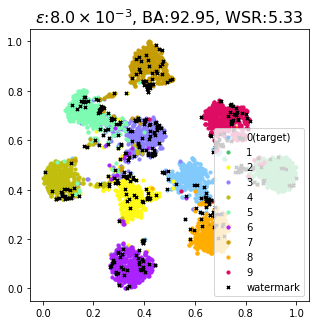}}
	\caption{t-SNE visualization of vanilla watermarked model along the adversarial direction.}
	\label{fig:tsne-adv-bd}
\end{figure*}

\begin{figure*}[!t]
	\centering
	\subfloat[]{\includegraphics[width=0.18\linewidth]{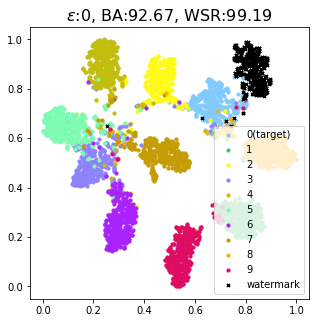}}
	\subfloat[]{\includegraphics[width=0.18\linewidth]{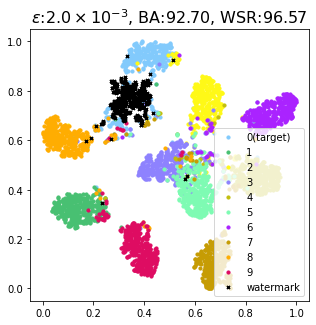}}
	\subfloat[]{\includegraphics[width=0.18\linewidth]{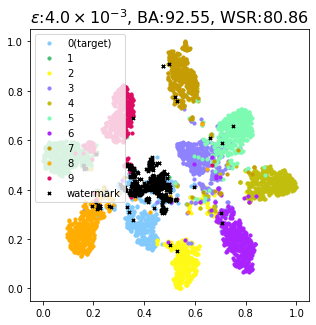}}
	\subfloat[]{\includegraphics[width=0.18\linewidth]{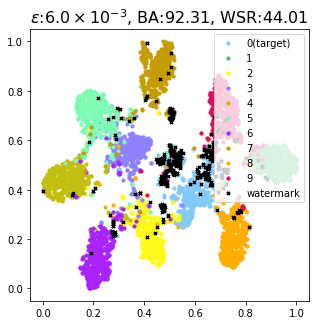}}
	\subfloat[]{\includegraphics[width=0.18\linewidth]{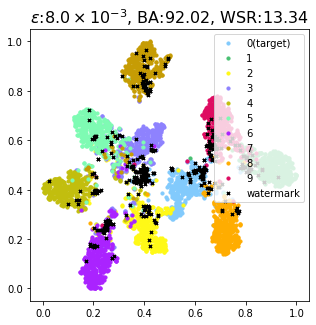}}
	\caption{t-SNE visualization of EW watermarked model along the adversarial direction.}
	\label{fig:tsne-adv-ew}
 \hspace{-2em}
\end{figure*}

\begin{figure*}[!htbp]
	\centering
	\subfloat[]{\includegraphics[width=0.18\linewidth]{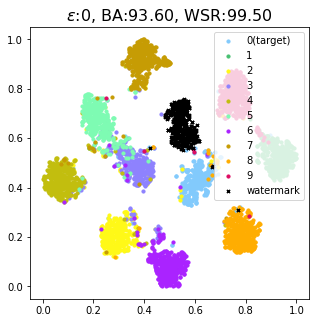}}
	\subfloat[]{\includegraphics[width=0.18\linewidth]{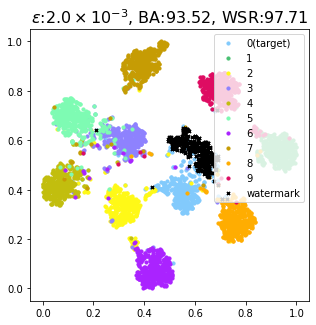}}
	\subfloat[]{\includegraphics[width=0.18\linewidth]{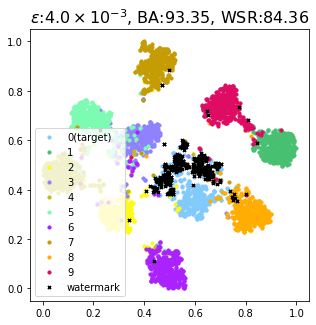}}
	\subfloat[]{\includegraphics[width=0.18\linewidth]{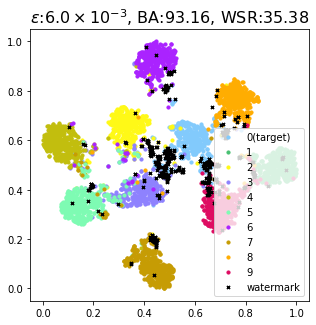}}
	\subfloat[]{\includegraphics[width=0.18\linewidth]{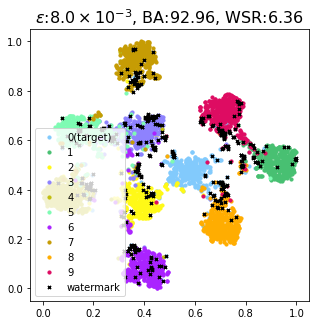}}
	\caption{t-SNE visualization of CW watermarked model along the adversarial direction.}
	\label{fig:tsne-adv-cw}
\end{figure*}

\begin{figure*}[!htbp]
	\centering
	\subfloat[]{\includegraphics[width=0.18\linewidth]{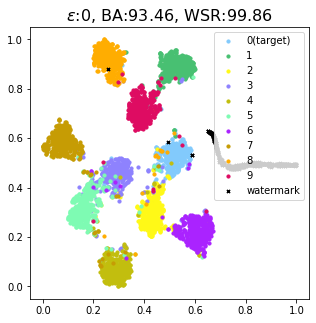}}
	\subfloat[]{\includegraphics[width=0.18\linewidth]{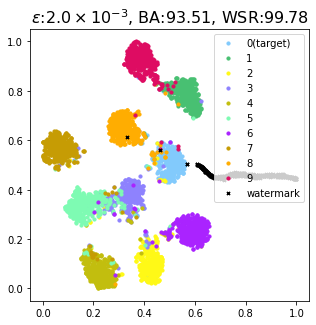}}
	\subfloat[]{\includegraphics[width=0.18\linewidth]{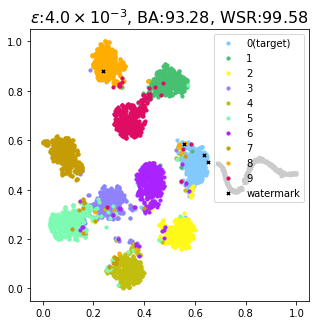}}
	\subfloat[]{\includegraphics[width=0.18\linewidth]{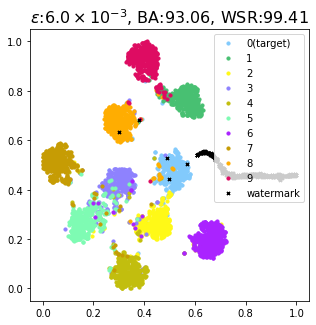}}
	\subfloat[]{\includegraphics[width=0.18\linewidth]{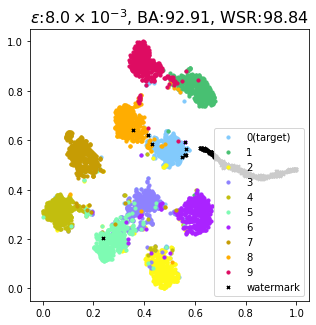}}
	\caption{t-SNE visualization of our watermarked model along the adversarial direction.}
	\label{fig:tsne-adv-ours}
% \hspace*{-1em}
\end{figure*}
%------------------------------------------------------------------------
\subsection{Feature Evolution During Fine-tuning}\label{app:tsne-ft}
We also investigate how the hidden representation evolves during the early stage of fine-tuning. We fine-tune the watermarked models for 200 iterations using the SGD optimizer with a learning rate of 0.05 and show how the representation evolves via t-SNE every 50 iterations. As can see in Figure~\ref{fig:tsne-ft-bd}-\ref{fig:tsne-ft-cer}, the representation of watermark samples quickly mixes with the clean representation in the early phase of fine-tuning, with the watermark success rate decreasing. While our method manages to maintain the watermark samples in a distinct cluster, and the cluster stays distant from the untargeted clusters during the fine-tuning process, as shown in Figure~\ref{fig:tsne-ft-ours}.
\begin{figure*}[!t]
	\centering
	\subfloat[]{\includegraphics[width=0.18\linewidth]{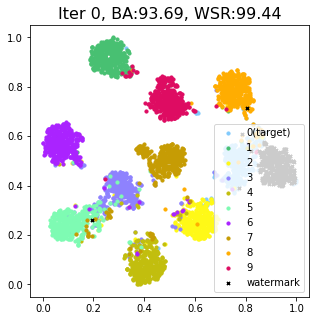}}
	\subfloat[]{\includegraphics[width=0.18\linewidth]{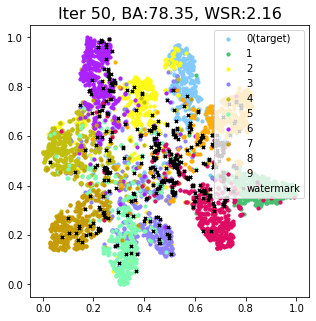}}
	\subfloat[]{\includegraphics[width=0.18\linewidth]{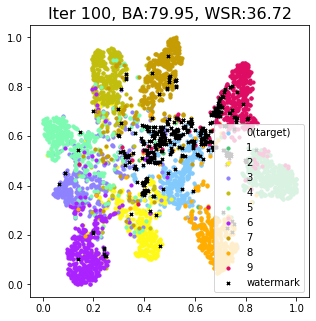}}
	\subfloat[]{\includegraphics[width=0.18\linewidth]{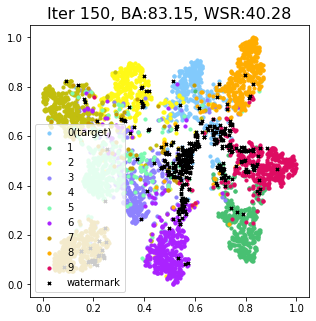}}
	\subfloat[]{\includegraphics[width=0.18\linewidth]{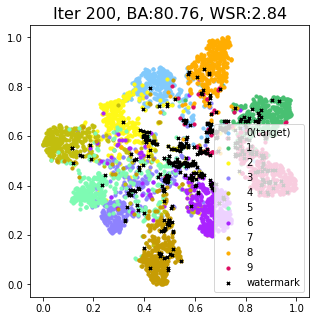}}
	\caption{t-SNE visualization of vanilla watermarked model during the process of fine-tuning.}
	\label{fig:tsne-ft-bd}
\end{figure*}
\begin{figure*}[!t]
	\centering
	\subfloat[]{\includegraphics[width=0.18\linewidth]{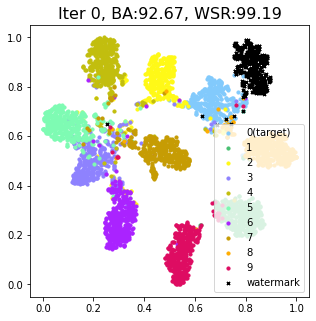}}
	\subfloat[]{\includegraphics[width=0.18\linewidth]{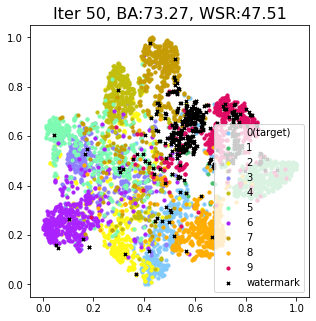}}
	\subfloat[]{\includegraphics[width=0.18\linewidth]{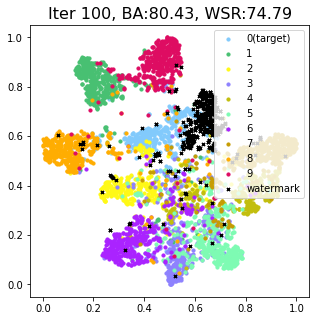}}
	\subfloat[]{\includegraphics[width=0.18\linewidth]{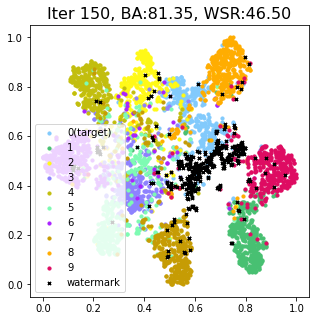}}
	\subfloat[]{\includegraphics[width=0.18\linewidth]{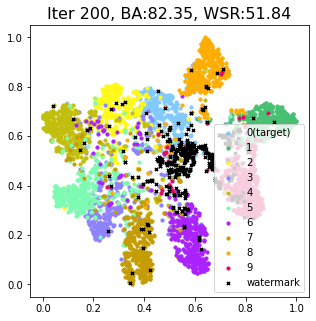}}
	\caption{t-SNE visualization of EW watermarked model during the process of fine-tuning.}
	\label{fig:tsne-ft-ew}
\end{figure*}
\begin{figure*}[!t]
	\centering
	\subfloat[]{\includegraphics[width=0.18\linewidth]{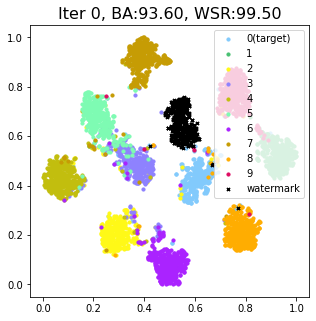}}
	\subfloat[]{\includegraphics[width=0.18\linewidth]{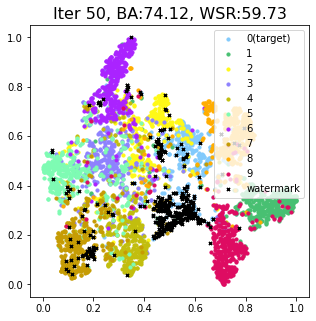}}
	\subfloat[]{\includegraphics[width=0.18\linewidth]{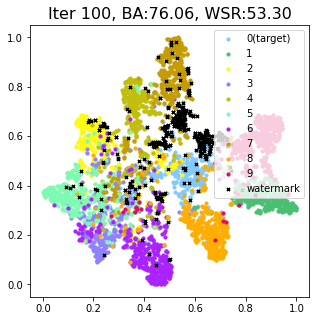}}
	\subfloat[]{\includegraphics[width=0.18\linewidth]{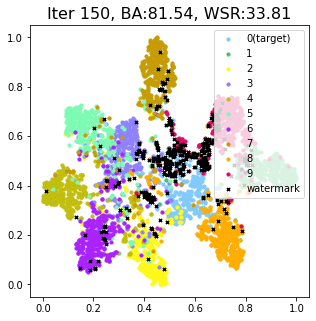}}
	\subfloat[]{\includegraphics[width=0.18\linewidth]{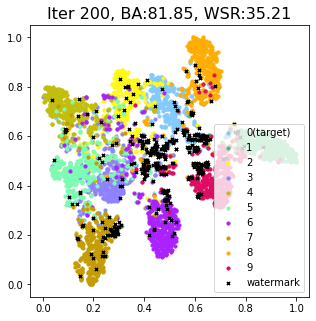}}
	\caption{t-SNE visualization of CW watermarked model during the process of fine-tuning.}
	\label{fig:tsne-ft-cer}
\end{figure*}
\begin{figure*}[!t]
	\centering
	\subfloat[]{\includegraphics[width=0.18\linewidth]{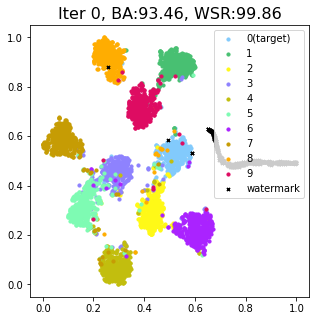}}
	\subfloat[]{\includegraphics[width=0.18\linewidth]{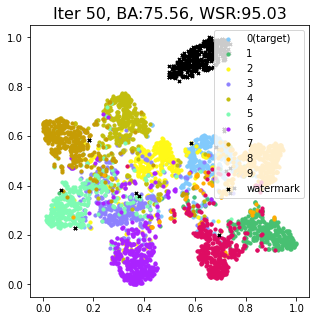}}
	\subfloat[]{\includegraphics[width=0.18\linewidth]{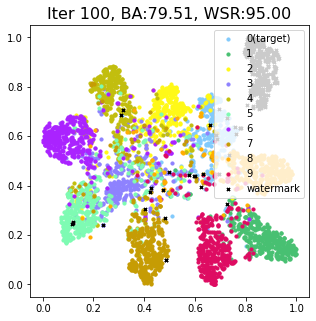}}
	\subfloat[]{\includegraphics[width=0.18\linewidth]{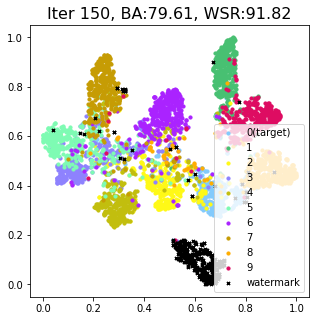}}
	\subfloat[]{\includegraphics[width=0.18\linewidth]{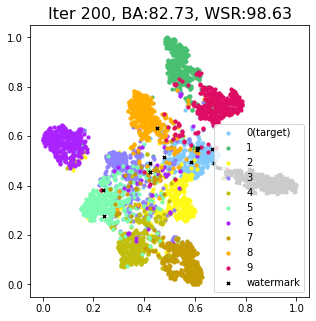}}
	\caption{t-SNE visualization of our watermarked model during the process of fine-tuning.}
	\label{fig:tsne-ft-ours}
\end{figure*}

%------------------------------------------------------------------------

% \include*{appendix}
% \input{appendix}

\end{document}